\documentclass[%
twocolumn,
superscriptaddress,
showpacs,
showkeys,
nofootinbib,
 amsmath,
 amssymb,
 aps,
 prl,
]{revtex4-1}

\usepackage{algorithm}
\usepackage{algorithmic}
\usepackage{graphicx}% Include figure files
\usepackage{dcolumn}% Align table columns on decimal point
\usepackage{bm}% bold math
\usepackage{amsthm} %theorem package
\usepackage{hyperref}% add hypertext capabilities
\usepackage[lofdepth,lotdepth]{subfig}
\usepackage{multirow}

\newcommand{\be}{\begin{equation}}
\newcommand{\ee}{\end{equation}}
\newcommand{\bea}{\begin{eqnarray}}
\newcommand{\eea}{\end{eqnarray}}
\newcommand{\mo}{\mod 1}
\newcommand{\smo}{_{\mod 1}}

\newtheorem{thm}{Theorem}

\newtheorem{Problem}{Problem}

\input{Qcircuit}

\begin{document}

%\preprint{APS/123-QED}

\title{Faster Phase Estimation}% Force line breaks with \\
%\thanks{A footnote to the article title}%

%\author{Matthew Hastings}
%\email{alexeib@microsoft.com}
% \altaffiliation[Also at ]{Physics Department, XYZ University.}%Lines break automatically or can be forced with \\
\author{Krysta M.~Svore}%
\email{ksvore@microsoft.com}
\affiliation{%
 Quantum Architectures and Computation Group\\
 Microsoft Research, Redmond, WA 98052 USA
}%
\author{Matthew B.~Hastings}%
\email{mahastin@microsoft.com}
\affiliation{%
 Station Q, Microsoft Research, Santa Barbara, CA 93106 USA
}%
\author{Michael Freedman}%
\email{michaelf@microsoft.com}
\affiliation{%
 Station Q, Microsoft Research, Santa Barbara, CA 93106 USA
}%

\date{\today}% It is always \today, today,
             %  but any date may be explicitly specified

\begin{abstract}
We develop several algorithms for performing quantum phase estimation based on basic measurements and classical post-processing.
We present a pedagogical review of quantum phase estimation and simulate the algorithm to numerically determine its scaling in circuit depth and width.
We show that the use of purely random measurements requires a number of measurements that is optimal up to constant factors, albeit at the cost of exponential classical post-processing; the method can also be used to improve classical signal processing.
We then develop a quantum algorithm for phase estimation that yields an asymptotic improvement in runtime, coming within a factor of $\log^*$ of the minimum number of measurements required while still requiring only minimal classical post-processing.
The corresponding quantum circuit requires asymptotically lower depth and width (number of qubits) than quantum phase estimation.
\end{abstract}

% PACS, the Physics and Astronomy
                             % Classification Scheme.
%Use showkeys class option if keyword
                              %display desired
\maketitle

%\tableofcontents

%\paragraph{\label{sec:Intro}Introduction---}

\section{Introduction}
Quantum algorithms promise computational speed-ups over their classical counterparts.
Quantum phase estimation is a key technique used in quantum algorithms, including algorithms for quantum chemistry \cite{AG2005,Chem2009} and quantum field theory \cite{Jordan11}, Shor's algorithm for prime factorization \cite{Shor1997}, and algorithms for quantum sampling \cite{TemmeEtAl11,ORR2012}.
It can be used to find eigenvalues of a unitary matrix efficiently.

There are two main approaches to quantum phase estimation: (1) invoking an inverse Quantum Fourier Transform (QFT) \cite{AL1999,Cleve1998,NielsenChuang2000} to extract information about the phase or (2) performing a basic measurement operation followed by classical post-processing in place of the QFT \cite{Kitaev1996,Kitaev2002}.
An advantage to approach (2) is that it uses classical post-processing in place of quantum operations, trading off an expensive resource for an inexpensive classical computation.
In particular, the QFT requires many small controlled-rotations, each of which must be approximated to precision $\epsilon$ by a sequence of basic quantum operations of length $O(\log(1/\epsilon))$ \cite{Selinger12}.
In practice, we may want to significantly reduce the circuit depth of the phase estimation algorithm in exchange for a small increase in circuit width, i.e., the number of qubits.
Therefore, we focus on approach (2) and rely primarily on quantum measurements to infer information about the phase.

We begin by outlining the goal of quantum phase estimation and explaining the basic measurement operation that is used as a subroutine to do this, and contrast this problem with the classical Fourier transform.
We then describe various phase estimation algorithms; these algorithms all call the same basic measurement operation, but use different parameters to do this.

We first present in Section \ref{sec:infthy} a technique based on random measurements to infer the phase;
this technique uses the fewest number of measurements of any we know (and we prove that it is within a constant factor of optimal),
but it requires impractical classical post-processing for use in, say, Shor's algorithm \cite{Shor1997},
with a complexity that is exponential in the number of bits being inferred.
However, this technique may be practical in certain classical noisy signal processing and inference applications, where the number of bits being inferred is smaller.
We explain these applications in this section and give some extensions of the technique that may be useful in inferring very noisy, sparse signals.

In Section \ref{sec:Kitaev}, we review a quantum phase estimation algorithm based on the same measurement operation, but the measurements are not random and the classical post-processing can be done efficiently \cite{Kitaev1996,Kitaev2002}.
We simulate this algorithm and determine its complexity, circuit depth, and circuit width for various sizes of input.

In Section \ref{sec:FastKitaev}, we improve upon this phase estimation algorithm by considering inference across multiple qubits.
We show that this technique requires asymptotically fewer measurements, and in turn has a correspondingly (asymptotically) smaller circuit width and depth, while still allowing efficient classical post-processing.

We compare the circuit constructions for Kitaev's phase estimation algorithm and the fast phase estimation algorithm in Section \ref{sec:circuits}.
Three models of computation are discussed: the first is a sequential model with limited parallelism, the second is a highly parallel model, and the third is a model based on a cluster of quantum computers.

\section{Phase Estimation and the Basic Measurement Operation}
We begin by reviewing the goal of quantum phase estimation and the basic measurement operator, following the algorithm of Kitaev \cite{Kitaev1996} (see Ref.~\onlinecite{Kitaev2002} for complete details).
We derive the steps slightly differently, in anticipation of our extension in the later sections.

Assume that we have a unitary operator $U$ and we would like to estimate the eigenvalues $\lambda_k$ of $U$ given $U$ and the eigenvectors $|\xi_k\rangle$:
\begin{equation}
U|\xi_k\rangle = \lambda_k|\xi_k\rangle,
\end{equation}
where the eigenvalues take the form $\lambda_k = e^{2\pi i\cdot \varphi_k}$.
The phase $\varphi_k$ is a real number modulo 1, which can be represented as a unit-length circle: $\varphi_k = \frac{k}{t} \mod 1$, $\varphi_k\in \mathbb{R}/\mathbb{Z}$, $0\leq k < t <2^m$ (while it may seem more natural at first to instead consider numbers that range between $0$ and $2\pi$, rather than choosing a number between $0$ and $1$ and multiplying by $2\pi$ as we do here, we choose the latter because it will be more natural when later considering an expansion of $\varphi_k$ as a binary fraction).
By measuring the eigenvalues of $U$, we can obtain an estimate
%, in the form of an irreducible fraction, of the phase $k/t$
of the phase $\varphi_k$; this process is called {\it quantum phase estimation}.
\footnote{In the context of Shor's algorithm, the corresponding eigenvector is defined as $|\xi_k\rangle = \frac{1}{\sqrt{t}}\sum_{n=0}^{t-1} e^{-2\pi i\cdot n\varphi_k} |a^n\rangle$.
%Each $a^n$ is an element of the cycle of $U$ containing $1$, i.e., $(1,a,a^2,\ldots,a^{t-1})$, where $t$ is the period of the cycle of $U$.
}

The goal of all phase estimation algorithms is to take a state of the form $|\xi_k\rangle$ and determine the corresponding eigenvalue $\lambda_k$.
The measurement operation described below commutes with $U$, so we can apply it multiple times to the same state with different parameters to improve our knowledge of the eigenvalue.
There are two parameters in the measurement result: (1) the {\it precision} $\delta$ and (2) the {\it probability of error} $\epsilon$.
That is, we obtain some estimate $\alpha$ of $\varphi_k$ where, with probability at least $1-\epsilon$, $|\alpha - \varphi_k|_{\mod 1} < \delta$, where $\mod 1$ is the distance on the unit circle.
If $\varphi_k$ is chosen from a discrete set of angles $\frac{k}{t}$ with fixed $t$ and unknown $k$, our goal is to make the precision smaller than $\delta = \frac{1}{2t}$ so that we can determine $\varphi_k$ exactly from this set.  In Section \ref{sec:infthy}, we slightly simplify the problem by directly inferring the angle $\varphi_k$ from the discrete set of possible angles, without bothering to introduce a real number precision; in this case $\epsilon$ is the probability of error in our discrete inference.

\subsection{Basic Measurement Operation}
We begin by constructing a measurement operator such that the conditional probability depends on $\varphi_k$, that is, upon measuring this operator, we learn some information about $\varphi_k$.
This construction relies on the fact that if $\ket{\xi_k}$ is an eigenvector of $U$, then it is also an eigenvector of powers $M$ of $U$:
\bea
U^M \ket{\xi_k} &=& \lambda_k^M \ket{\xi_k} \\ \nonumber
%&=& (e^{2\pi i \cdot \varphi_k})^M\ket{\xi_k} \\ \nonumber
&=& e^{2\pi i M \cdot \varphi_k} \ket{\xi_k}. \nonumber
\eea

The operator takes as input two quantum registers: one initialized to $|0\rangle$ and the other initialized to the eigenvector $|\xi_k\rangle$.
%In fact, the second quantum register actually contains $\ket{1} = \frac{1}{\sqrt{t}}\sum_{k=0}^{t-1}\ket{\xi_k}$.
The operator depends upon two parameters, a ``multiple" $M$ and an ``angle" $\theta$, where $M$ is an integer between $1$ and $t-1$ (to make it practical to implement, we restrict to positive integers $M$) and $\theta$ is a real number between $0$ and $2\pi$.

\begin{figure}
\centering
\[
\Qcircuit @C=1em @R=1em {
\lstick{\ket{0}} & \gate{H}  & \gate{Z(\theta)}   &\ctrl{1}   & \gate{H}    & \meter    & \cw \\
\lstick{\ket{\xi_k}} &   {/} \qw   &\qw   & \gate{U^M} &    \qw      &    \qw    & \qw }
\]
\caption{Circuit to perform the measurement operator.}
\label{fig:measCircuit}
\end{figure}
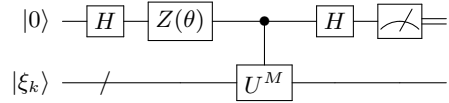

The measurement operator used to measure the eigenvalues is as follows:
\bea
&& \Xi_{M,\theta}(U) \\ \nonumber
%= \sum_k V_k \otimes \prod_{\mathcal{L}_k} V_k
&= &
\sum_k \frac{1}{2}
\left[ \begin{array}{ c c }
     1+e^{2\pi i M\cdot\varphi_k+i\theta} & 1-e^{2\pi i M\cdot\varphi_k+i\theta} \\
     1-e^{2\pi iM\cdot\varphi_k+i\theta} & 1+e^{2\pi i M\cdot\varphi_k+i\theta}
  \end{array} \right] \otimes |\xi_k\rangle \langle \xi_k| \\ \nonumber
&=& \frac{1}{2}\left[\begin{array}{ c c}
1+U^M\exp(i\theta) & 1-U^M \exp(i\theta) \\
1-U^M \exp(i\theta) & 1+U^M \exp(i\theta)
\end{array}\right],
\eea
which acts on the quantum states by the following transformation:
\bea
&&
|0\rangle\otimes \ket{\xi_k}
%\left[\ket{1} = \sum_k|\xi_k\rangle\right]
 \overset{\Xi_{M,\theta}(U)}
{\mapsto}
 \\ \nonumber
&&  \left(\frac{1+e^{2\pi i M\cdot \varphi_k+i\theta}}{2}|0\rangle + \frac{1-e^{2\pi iM\cdot \varphi_k+i\theta}}{2}|1\rangle\right)\otimes|\xi_k\rangle.
\eea
The corresponding circuit is shown in Fig.~\ref{fig:measCircuit}.  The gate $Z(\theta)$ corresponds to the unitary:
\be
Z(\theta) =
\left[\begin{array}{ c c}
1 & 0 \\
0 & e^{i\theta}
\end{array}\right].
\ee

It follows that the measurement outcome probabilities are given by:
\begin{equation}
\label{cmp1}
P_{M,\theta}(0|k) = \left| \frac{1+e^{2\pi i M\cdot \varphi_k+i\theta}}{2} \right|^2 = \frac{1 + \cos(2\pi M \cdot \varphi_k + \theta)}{2},
\end{equation}
and
\begin{equation}
\label{cmp2}
P_{M,\theta}(1|k) = \left| \frac{1-e^{2\pi i M \cdot\varphi_k+i\theta}}{2} \right|^2 = \frac{1 - \cos(2\pi M \cdot \varphi_k + \theta)}{2}.
\end{equation}
We write these probabilities as conditional probabilities to emphasize that they depend upon the unknown $k$.

\subsection{Relation To Classical Fourier Transform and Generalizations}
\label{sec:CFT}
With Eqs.~(\ref{cmp1},\ref{cmp2}) in hand, we can see that if we apply a large number of measurements using the same $M$ at both $\theta=0$ and $\theta=\pi/2$, we can accurately estimate $\cos(2 \pi M \cdot\varphi_k)$ and $\sin(2 \pi M \cdot\varphi_k)$.  Using a sufficiently accurate estimate of these cosines and sines at two different values of $M$ allows us to determine $\varphi_k$ accurately.  This is the problem of reconstructing a sparse signal (in this case, composed of a single Fourier mode) from its value at a small number of different ``times" (i.e., different values of $M$).  However, the accurate determination of $\cos(2\pi M\cdot\varphi_k)$ would require a very large number of measurements, polynomial in $t$, while other methods require many fewer measurements.  The reason is that the large number of measurements at a fixed value of $M$ means that each measurement imparts little additional information.  By varying $M$, we are able to obtain accurate results from a much smaller number of measurements.

This relates to a problem of reconstructing the Fourier transform of a signal from very noisy measurements.  The quantum phase estimation problem involves a signal with a single Fourier mode.  However, this gives rise to a natural generalization of reconstructing a problem with a small number of Fourier modes from very noisy measurements.  We consider this problem at the end of the next section.

\section{``Information Theory" Phase Estimation}
\label{sec:infthy}
One procedure for estimating the phase (or angle) is to perform a series of random measurements and then solve a hard classical reconstruction problem.
We measure the operator at a set of randomly chosen multiples $M_i$ and angles $\theta_i$ and classically reconstruct the angle $2\pi \varphi_k$.  In this section, we show that we can determine $\varphi_k$ with only $O(\log(t))$ measurements; we also show that this result is tight.

We randomly select $M_i$ for each measurement $i$ between $1$ and $t-1$, and also assume a small randomized offset noise $\theta_i = 2\pi r$, where $r$ is a random double.
The conditional measurement probabilities for this measurement operator on the $i^{\textrm{th}}$ measurement are given by:
\begin{equation}
\label{cmpx1}
P_i(0|k) = \frac{1 + \cos(2 \pi M_i \cdot \varphi_k + \theta_i)}{2},
% = \cos^2(\theta_i \cdot 2\pi \varphi_k + theta_i),
\end{equation}
and
\begin{equation}
\label{cmpx2}
P_i(1|k) = \frac{1 - \cos(2\pi M_i \cdot \varphi_k + \theta_i)}{2} = 1 - P_i(0|k).
\end{equation}

Let $v_i$ be the outcome of the $i^{\textrm{th}}$ measurement.
Since different measurements are independent events, the probability of getting a given sequence of measurement outcomes is
\begin{equation}
\label{maxLikelihood}
P(v_1,\ldots,v_s|k) = \prod_{i=1}^s P_i(v_i|k).
\end{equation}
Assuming a flat a priori distribution of $k$, the probability distribution of $k$ given the measurement sequence is proportional to $P(v_1,\ldots,v_s|k)$.
The algorithm then to compute $k$ given a sequence of $s$ measurements is simple: for each $k$ compute the probability $P(v_1,\ldots,v_s|k)$, outputting the $k$ which maximizes this.  The post-processing time required is of order $s t$, which is exponentially large in the number of bits inferred since the value of $k$ that it outputs can be written with $\lceil \log_2(t) \rceil$ bits.

The information theory phase estimation algorithm is given in Algorithm \ref{alg:InfTheory}.
\begin{algorithm}[H]
\caption{Information Theory Phase Estimation}
\label{alg:InfTheory}
\algsetup{indent=2em}
\begin{algorithmic}[1]
\FOR{$i=1$ to $s$}
\STATE{Choose random $M_i$.  Choose random $\theta_i$.
}
\STATE{Perform basic measurement operation with multiple $M_i$ and angle $\theta_i$.}
\label{line:inferinf}
\ENDFOR
\STATE{Maximize
\be
P(v_1,\ldots,v_s|k) = \prod_{i=1}^s P_i(v_i|k) \nonumber
\ee
%\end{equation}
over all choices of $k$.}
\RETURN{$k/t$, the estimate of the phase.}
\end{algorithmic}
\end{algorithm}

\begin{figure*}[tb]
\centering
\subfloat[][10 random measurements.]{
\includegraphics[width=0.45\textwidth]{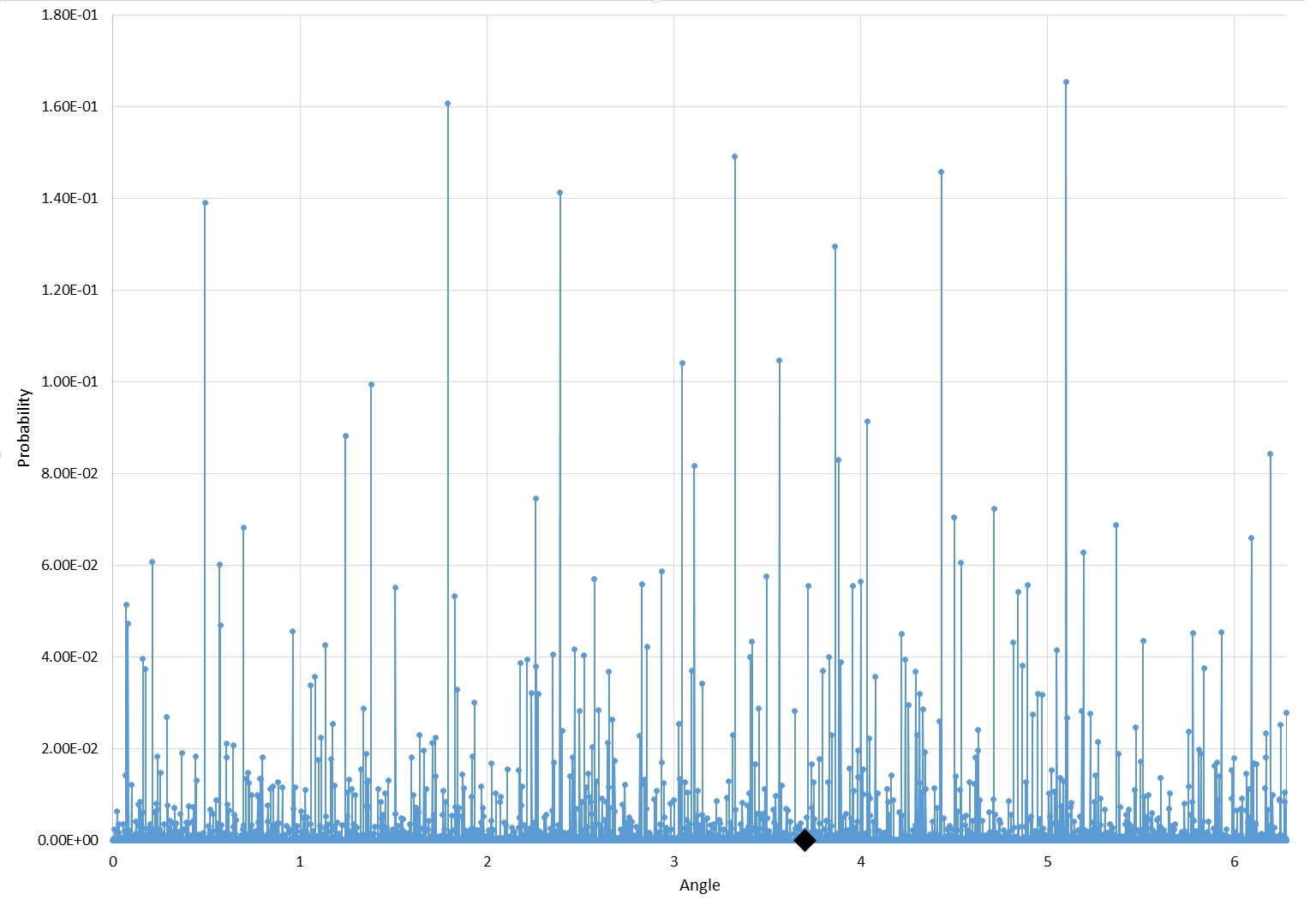}
  \label{fig:InfTheory10}
}
\qquad
\subfloat[][20 random measurements.]{
\includegraphics[width=0.45\textwidth]{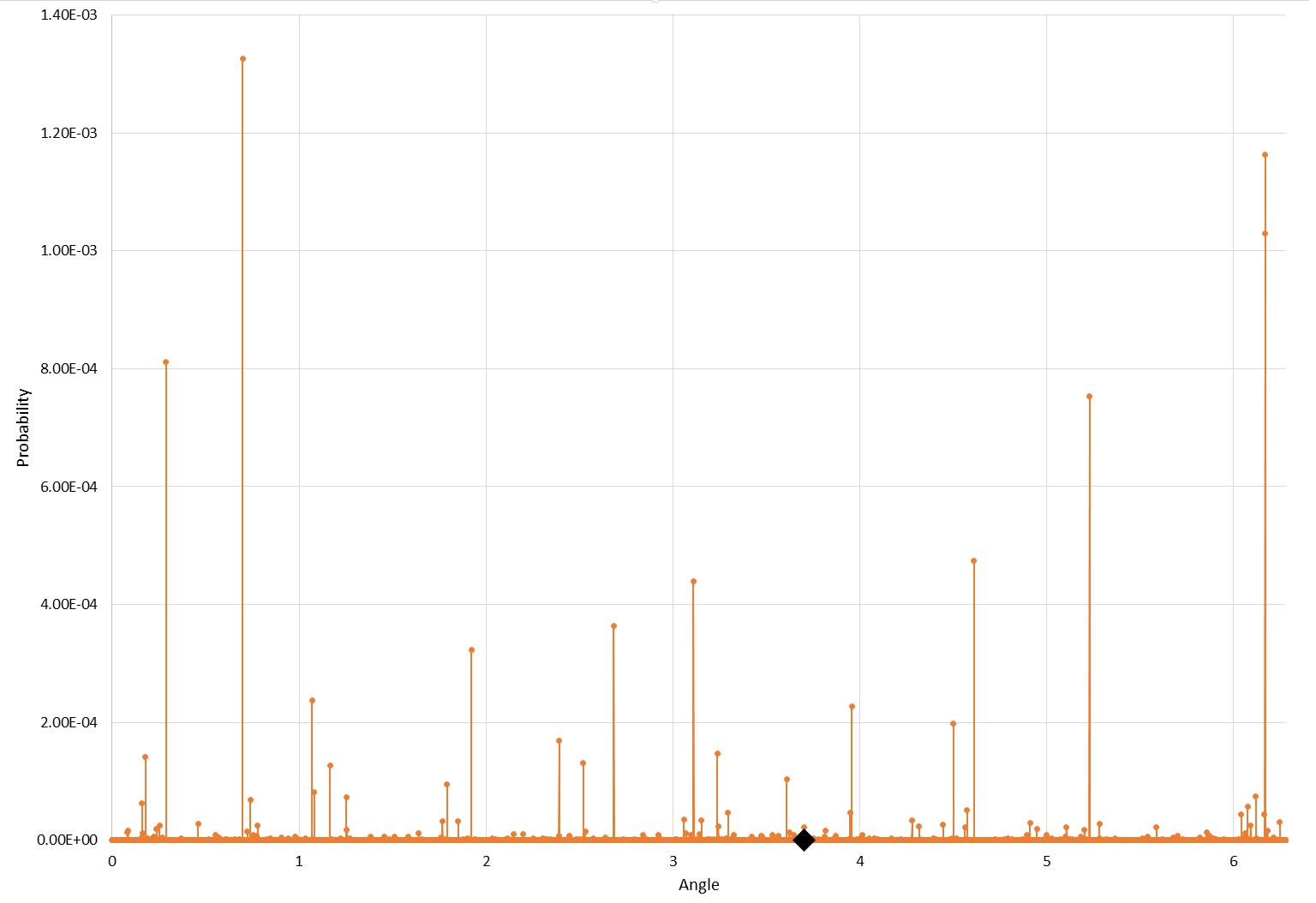}
  \label{fig:InfTheory20}
}
\subfloat[][30 random measurements.]{
\includegraphics[width=0.45\textwidth]{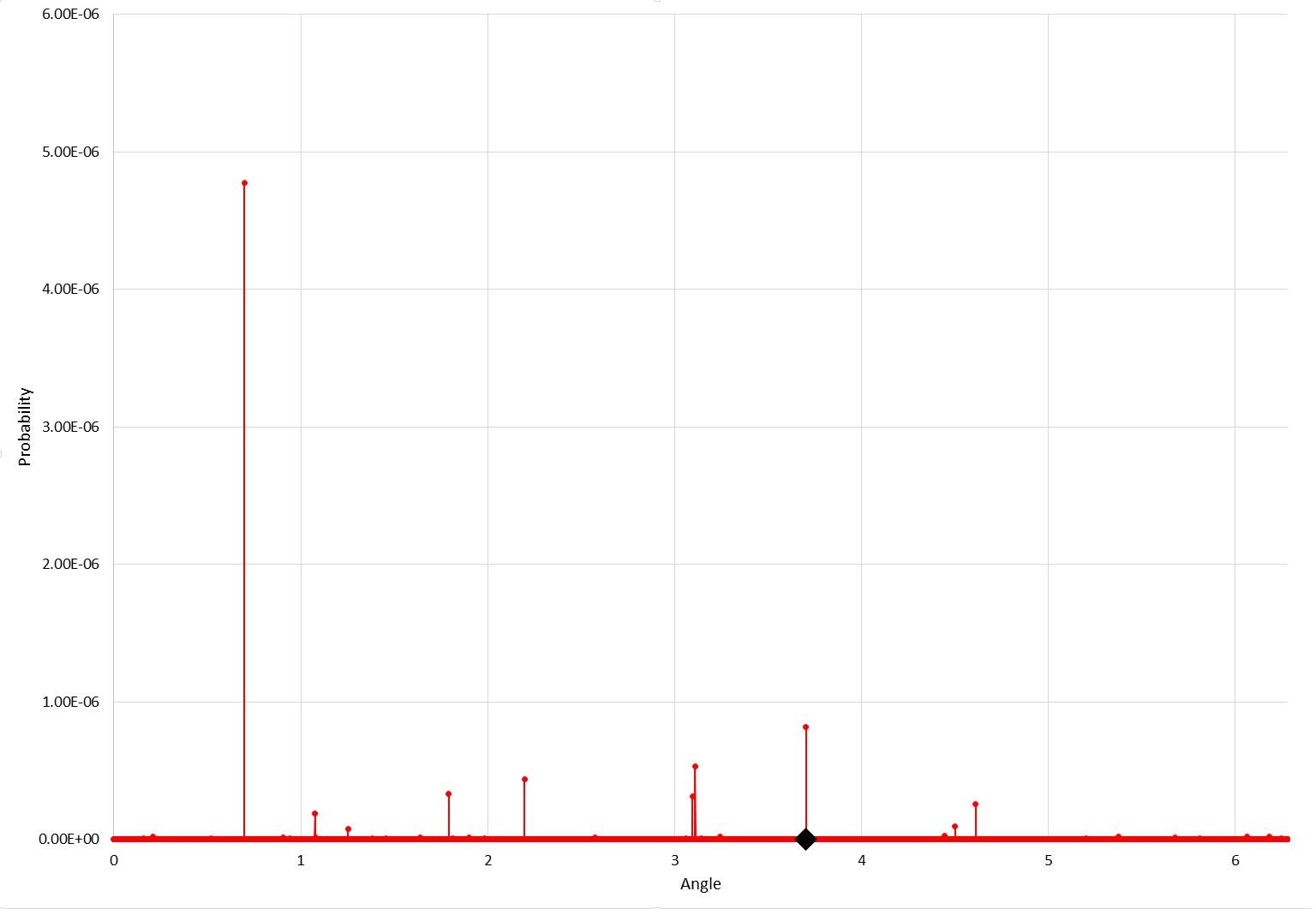}
  \label{fig:InfTheory30}
}
\qquad
\subfloat[][40 random measurements.]{
\includegraphics[width=0.45\textwidth]{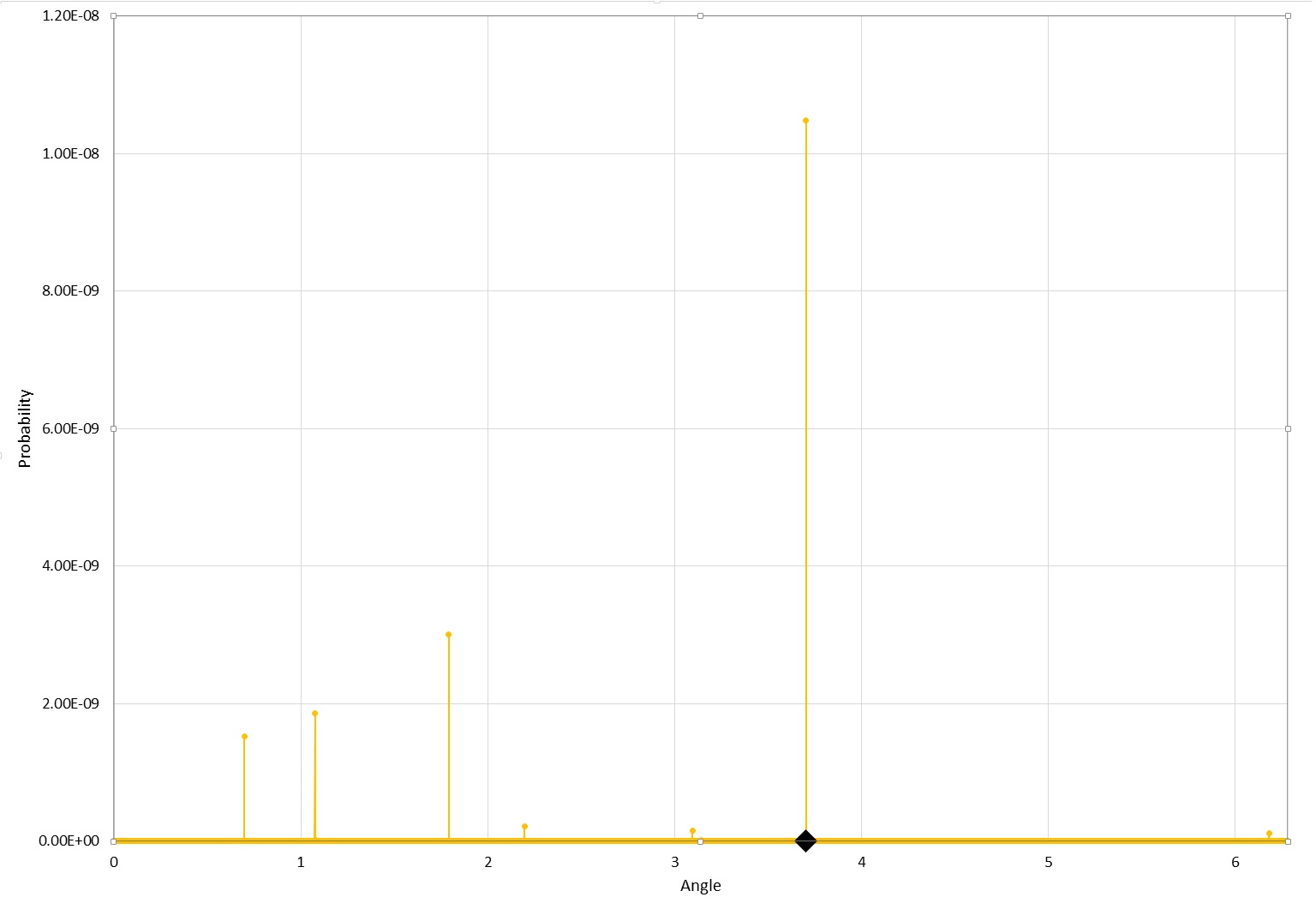}
  \label{fig:InfTheory40}
}
\subfloat[][50 random measurements.]{
\includegraphics[width=0.45\textwidth]{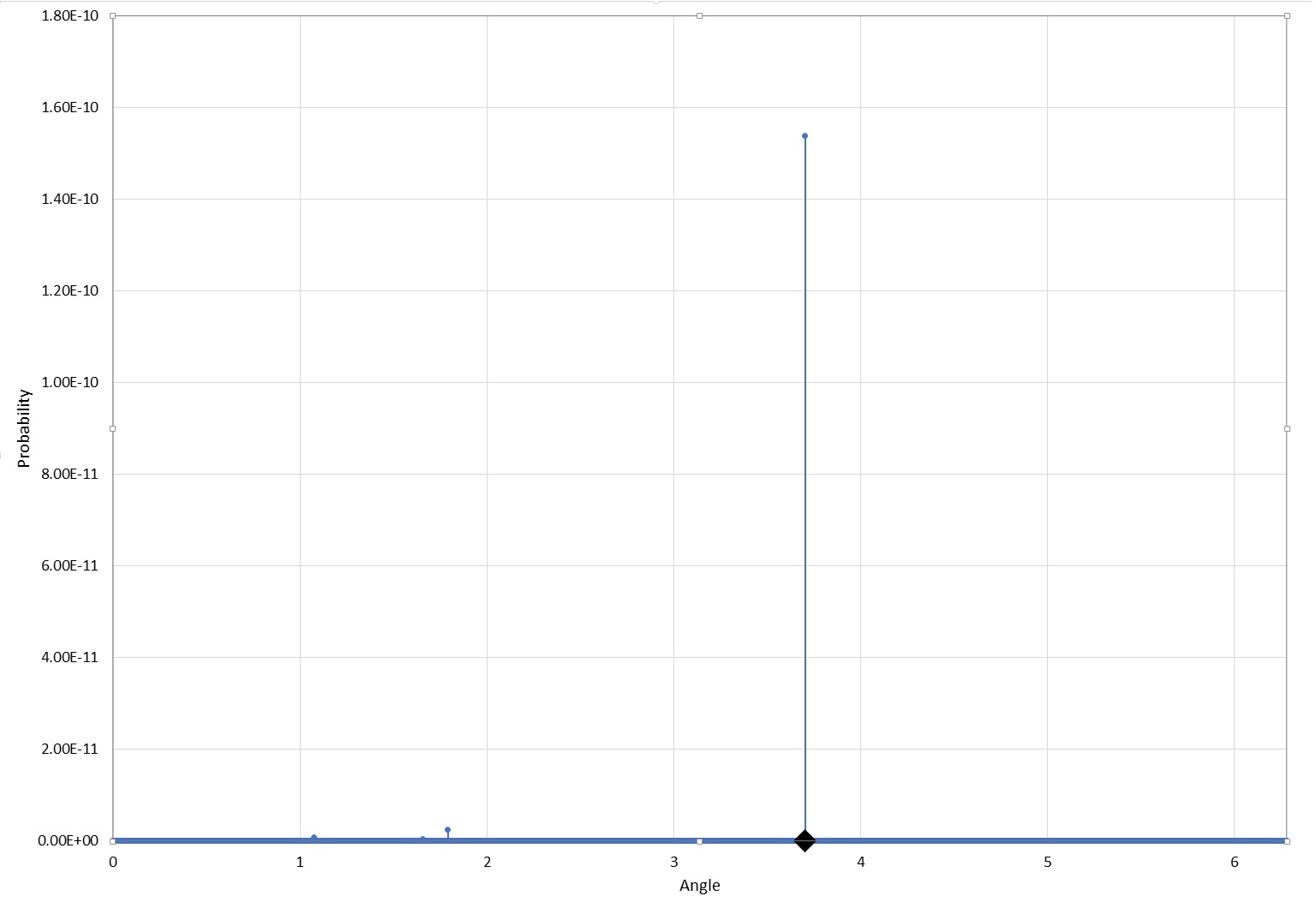}
  \label{fig:InfTheory50}
}
\caption{Results of simulating the probability of inferring a given angle among $t = 10000$ equally distributed possible angles.
Plots are of the inferred probability distribution as a function of angle after 10--50 random measurements.
The correct angle is marked by a black diamond.
}
\end{figure*}

To illustrate,
we simulated the probability of inferring the given angle $2\pi \varphi_k$ among $t=10^4$ equally distributed possible angles.
Figures \ref{fig:InfTheory10}--\ref{fig:InfTheory50} plot the inferred probability distribution as a function of angle after $s$ measurements, where $s=\{10,20,30,40,50\}$.
The black diamond on each plot indicates the peak at the correct angle.
From the plots, we see that after 10--20 measurements, the inference is very noisy, while after 40 or more measurements it has inferred {\it some} information about the correct angle, and after 50 measurements it is very precise.

%\begin{figure}[htb]
%  \centering
%  \includegraphics[width=3.5in]{t30.pdf}
%  \caption{Hard classical reconstruction: Inferred angle (correct is at 0) vs.~ inferred probability of angle.  Black: 10, Red: 20, Green: 30 measurements, respectively.  Y-axis currently arbitrary (shifted for appearance).  100 possible outcomes.}
%  \label{fig:InfTheory}
%\end{figure}

%\begin{figure}[htb]
%  \centering
%  \includegraphics[width=3.5in]{10s.pdf}
%  \caption{Hard classical reconstruction: Inferred angle vs.~ inferred probability of angle.  10000 possible outcomes.  10 measurements of the random angle.  Black diamond marks correct angle.}
%  \label{fig:InfTheory10}
%\end{figure}

%\begin{figure}[htb]
%  \centering
%  \includegraphics[width=3.5in]{20s.pdf}
%  \caption{Hard classical reconstruction: Inferred angle vs.~ inferred probability of angle.  10000 possible outcomes.  20 measurements of the random angle.  Black diamond marks correct angle.}
%  \label{fig:InfTheory20}
%\end{figure}

%\begin{figure}[htb]
%  \centering
%  \includegraphics[width=3.5in]{30s.pdf}
%  \caption{Hard classical reconstruction: Inferred angle vs.~ inferred probability of angle.  10000 possible outcomes.  30 measurements of the random angle. Black diamond marks correct angle.}
%  \label{fig:InfTheory30}
%\end{figure}

%\begin{figure}[htb]
%  \centering
%  \includegraphics[width=3.5in]{40s.pdf}
%  \caption{Hard classical reconstruction: Inferred angle vs.~ inferred probability of angle.  10000 possible outcomes.  40 measurements of the random angle. Black diamond marks correct angle.}
%  \label{fig:InfTheory40}
%\end{figure}

%\begin{figure}[htb]
%  \centering
%  \includegraphics[width=3.5in]{50s.pdf}
%  \caption{Hard classical reconstruction: Inferred angle vs.~ inferred probability of angle.  10000 possible outcomes.  50 measurements of the random angle. Black diamond marks correct angle.}
%  \label{fig:InfTheory50}
%\end{figure}

\begin{figure*}[tb]
\centering
\includegraphics[width=6in]{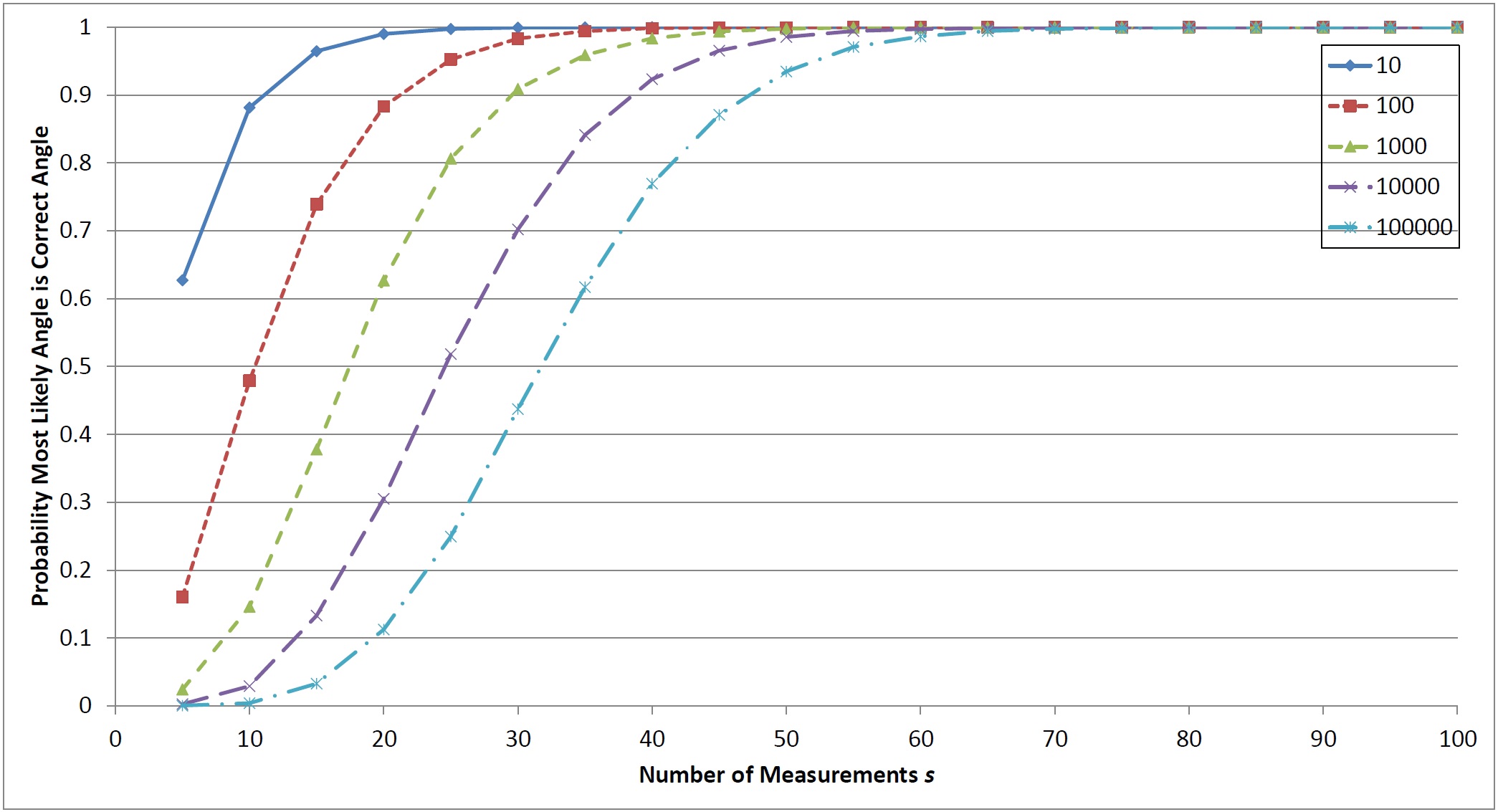}
\caption{Results of simulating ``information theory" phase estimation for $t = \{10^1,10^2,10^3,10^4,10^5\}$ equally distributed possible angles.
The $x$-axis is the number of random measurements $s$; the $y$-axis is the probability that the most likely angle after $s$ measurements is the correct angle.
}
\label{fig:InfTheoryOverview}
\end{figure*}

In Figure \ref{fig:InfTheoryOverview}, we plot simulation results for inferring the given angle $2\pi\varphi_k$ among $t = \{10^1,10^2,10^3,10^4,10^5\}$ equally distributed possible angles.
The $x$-axis is the number of random measurements $s$ and the $y$-axis is the probability that the $k$ which maximizes Eq.~(\ref{maxLikelihood}) after $s$ measurements is the correct angle.
Clearly, as $t$ increases, the number of measurements increases, following an $O(\log(t))$ behavior.

%\begin{figure*}
%\centering
%\includegraphics[width=6in]{SummaryAngles.jpg}
%\caption{Results of simulating ``information theory" phase estimation for $t = \{10^2,10^3,10^4,10^5,10^6\}$ equally distributed possible angles.
%Plots are of the inferred probability of the correct angle as a function of conducting $s$ random measurements.}
%\label{fig:InfTheoryOverview}
%\end{figure*}

\subsection{Bounds on $s$}
We now show that $O(\log(t))$ measurements suffice to estimate the angle with high probability.
This number of measurements required is asymptotically optimal (up to constant factors), as
clearly $\lfloor \log_2(t) \rfloor$ measurements are required to have an error probability greater than $1/2$: after $s$ measurements, there are at most $2^s$ possible outcomes for the sequence of measurements, so to select an angle from a set of $t$ choices with probability greater than $1/2$, we need $2^s>t/2$.  A more sophisticated entropic argument would likely be able to improve the constant in front of this lower bound.

The next theorem implies that the number of measurements to obtain error probability at most $\epsilon$ is $\log_{1/c}(t/\epsilon)$ for some constant $c<1$.
\begin{thm}
Suppose we choose the multiples $M_i$ and angles $\theta_i$ at random as above.
Suppose the measurement outcomes are chosen with probabilities given in Eqs.~(\ref{cmpx1},\ref{cmpx2}) for $k=k_0$.  Then,
the probability $\epsilon$ that the algorithm described above chooses a $k' \neq k_0$ as the choice with maximal likelihood is bounded by
\be
t c^{s}
\ee
for some numerical constant $c$ strictly less than $1$ ($c$ does not depend upon $t$).
\begin{proof}
We first consider a given $k' \neq k_0$ and estimate the probability that after $s$ measurements, the probability
$P(v_1,\ldots,v_s|k') = \prod_{i=1}^s P_i(v_i|k')$ is greater than or equal to $P(v_1,...v_s|k_0)$.
Consider the expectation value
\be
\label{consex}
E\Bigl[\frac{P(v_1,\ldots,v_s|k')^{1/2}}{P(v_1,\ldots,v_s|k_0)^{1/2}}\Bigr],
\ee
where the expectation value is over measurement outcomes and choices of $M_i$ and $\theta_i$.
This equals
\bea
\nonumber
&& E_{\{M_i,\theta_i\}}\Bigl[ \sum_{\{v_i\}} \frac{P(v_1,\ldots,v_s|k')^{1/2}}{P(v_1,\ldots,v_s|k_0)^{1/2}} P(v_1\ldots,v_s|k_0)\Bigr]
\\ \nonumber
&=&  E_{\{M_i,\theta_i\}}\Bigl[\sum_{\{v_i\}} P(v_1,\ldots,v_s|k')^{1/2} P(v_1,\ldots,v_s|k_0)^{1/2}\Bigr],
\eea
where the sum is over all $2^s$ possible sequences $v_1,...,v_s$ of measurement outcomes and the expectation value is now over all choices of $\theta_i,M_i$.
This equals
\be
\Bigl( E_{M,\theta} \Bigl[\sum_v P_{M,\theta}(v|k')^{1/2} P_{M,\theta} (v|k_0)^{1/2} \Bigr] \Bigr)^s,
\ee
where $E_{M,\theta}[...]$ is the expectation value over $M,\theta$.
A direct calculation shows that for all $k'\neq k$, the term in parenthesis  $E_{M,\theta}[\sum_v P_{M,\theta} (v|k')^{1/2} P_{M,\theta} (v|k_0)^{1/2}]$ is bounded by some constant $c<1$ for all $t$.
Thus, the expectation value (\ref{consex}) is bounded by $c^s$.
Thus, for a given $k'$, the probability that $P(v_1,\ldots,v_s|k') \geq P(v_1,\ldots,v_s|k_0)$ is bounded by $c^{s}$, as can be shown by applying Markov's inequality to $\frac{P(v_1,\ldots,v_s|k')^{1/2}}{P(v_1,\ldots,v_s|k_0)^{1/2}}$.

Thus, the probability that there is a $k'$ such that $P(v_1,\ldots,v_s|k') \geq P(v_1,\ldots,v_s|k_0)$ is bounded by $t c^{s}$.
\end{proof}
\end{thm}
We have not bothered to optimize the estimate in the above theorem: it is possible that a tighter bound could be considered by estimating the expectation value
$E[\Bigl(\frac{P(v_1,\ldots,v_s|k')}{P(v_1,\ldots,v_s|k_0)}\Bigr)^a]$ for some constant $0<a<1$ and optimizing the choice of $a$ in the spirit of the Chernoff bound.

Finally, we remark that while we have selected $\theta$ randomly between $0$ and $2\pi$ in the above algorithm and in the above theorem, in fact it would suffice to pick $\theta$ randomly from the set of angles $\{0,\pi/2\}$, or indeed from any set of a pair of angles that do not differ by exactly $\pi$ (for example, the set $\{0,\pi\}$ would not work).  The proof of the theorem would be essentially the same in this case, and restricting to such a smaller set of angles may be more convenient for implementation on a quantum computer.

\subsection{Classical Inference of Multiple Fourier Modes}
The results above suggest a natural generalization of the problem.  Define a classical channel $E(x)$ which maps from a real number $x$ between $-1$ and $1$ to an output consisting of a single bit.
We fix the output probabilities of this channel:
\be
\label{Pc0}
P(0|x)=\frac{1+x}{2},
\ee
\be
\label{Pc1}
P(1|x)=\frac{1-x}{2}.
\ee
Then Eqs.~(\ref{cmpx1},\ref{cmpx2}) can be interpreted as follows: for $\theta_i=0$, for any $M_i$, we take the number $\cos(2 \pi M_i \cdot \varphi_k)$ and input this number into the channel and the output of the channel is the measurement outcome, while for $\theta_i=1$, we instead input $\sin(2 \pi M_i \cdot \varphi_k)$.

This then suggests a natural generalization.  Consider a classical signal written as a sum of Fourier modes:
\be
f(M)=\sum_k a(k) \exp(2 \pi i M \cdot \varphi_k).
\ee
Here, $M$ is an integer and the function is periodic with period $t$.

Then, we have the natural classical problem:
\begin{Problem}
Assume that $f(M)$ is $K$-sparse, meaning that at most $K$ of the coefficients $a(k)$ are non-zero.
Assume that the non-zero $a(k)$ are chosen from a discrete set $S$ of possible values (typically we will be interested in $|S|$ being small), with
${\rm min}_{a \neq b, a\in S, b\in S} |a-b| \geq d_{min}$ for some $d_{min}$.  The $a(k)$ may be complex.

Let $A_{max}$ be the maximum of $|f(M)|$ over all such $K$-sparse $a(k)$ and over all $M$.

Assume that we have some channel $C(x)$ which maps from a real number in the range $[-A_{max},A_{max}]$ to an output chosen from a discrete set (the channel $C(x)$ need not be the same as that given above in Eqs.~(\ref{Pc0},\ref{Pc1})).  For this channel to be useful in inferring $x$ from measurements of the output, we will require that different input numbers lead to different output probabilities, and we will quantify this more precisely below in Eq.~(\ref{c0assumption}).

Pick several different $M_i$, and for each $M_i$
measure $C({\rm Re}(f(M_i)))$ or $C({\rm Im}(f(M_i)))$.  Infer the coefficients $a(k)$.
\end{Problem}

This problem can be interpreted as inferring a classical sparse signal from noisy measurements at several different ``times" (interpreting each $M_i$ as a time at which to infer the signal).
We now show, given suitable assumptions on $C(x)$, that this problem can be solved using a number of measurements that is $O(\log(N_{choices}))$, where $N_{choices}$ is the number of possible choices of $K$-sparse $f(M)$.
As before, this number of measurements is asymptotically optimal. Note that for $K<<t$, $\log(N_{choices}) \approx K\log(t|S|)$.

The procedure we describe is similar to that previously: we select random $M_i$ and randomly choose whether to measure $C({\rm Re}(f(M_i)))$ or $C({\rm Im}(f(M_i)))$ at each time.
After $s$ measurements, we select the choice of $a(k)$ which has the maximal a posteriori likelihood, assuming a flat initial distribution.
Interestingly, since the number of measurements we need is asymptotically much smaller than $\sqrt{t}$ (indeed we only need $O(\log(t))$ measurements if $K=O(1)$), this means that this random procedure typically does not ever pick $M_i=M_j$ for $i\neq j$.  That is, interpreting the $M_i$ as ``times", this means that we do not ever measure the signal twice at the same time.

Note that we have assumed that the non-zero coefficients are chosen from a small set $S$ of possible values.  As the number of possible values of $S$ increases, the number of measurements increases for two reasons.  First $N_{choices}$ increases.  Second, the values in the set become more closely spaced ($d_{min}$ becomes smaller compared to $A_{max}$), and the measurement outcomes probabilities hence become less sensitive to the particular value of $a(k)$.  This second problem is actually the more serious one.  Suppose that we have a signal that is $1$-sparse, and we even know that the only non-zero $a_k$ is at $k=0$.  The question is to infer the magnitude of $a_0$.  Every measurement then consists of sending $a_0$ into the channel $C(x)$.  Using the channel $C(x)$ before, it takes $1/\epsilon^2$ measurements to infer $a_0$ to precision $\epsilon$.  This number of measurements is exponential in the number of bits of precision in $a_0$.  That is, it takes many more measurements to infer the amplitude of a Fourier coefficient than it does to infer its frequency.

\begin{thm}
Suppose we choose the multiples $M_i$ at random as above and randomly choose whether to measure $C({\rm Re}(f(M_i)))$ or $C({\rm Im}(f(M_i)))$ at each time.
Suppose also that
$C(x)$ has the probability that for any $x,y \in [-A_{max},A_{max}]$ we have
\be
\label{c0assumption}
\sum_v P(v|x)^{1/2} P(v|y)^{1/2} \leq 1-c_0 |x-y|^2
\ee
for some constant $c_0$,
where the probabilities $P(v|x)$ are the probability that the channel $C$ gives output $v$ given input $x$.
Then,
the probability $\epsilon$ that the algorithm described above chooses a $k' \neq k_0$ as the choice with maximal likelihood is bounded by
\be
N_{choices} c^{s}
\ee
where
\bea
c &\leq &1-c_0 \Bigl(\frac{d_{min}}{2} \Bigr)^2 \frac{d_{min}^2}{16 A_{max}^2}.
\eea
\begin{proof}
Assume the correct choice of $a(k)$ is given by $a_0(k)$.  We consider a given sequence $a'(k)$ (such that for at least one $k$, $a'(k) \neq a_0(k)$) and
 estimate the probability that after $s$ measurements, the probability
$P(v_1,\ldots,v_s|a'(k))$ is greater than or equal to $P(v_1,...v_s|a_0(k))$, where $v_1,\ldots,v_s$ are the measurement outcomes of the channel.

Let
\be
f_0(M)=\sum_k a_0(k) \exp(2 \pi i M \cdot \varphi_k)
\ee
and
\be
f'(M)=\sum_k a'(k) \exp(2 \pi i M \cdot \varphi_k).
\ee

Consider the expectation value
\be
\label{consexgen}
E\Bigl[\frac{P(v_1,\ldots,v_s|a'(k))^{1/2}}{P(v_1,\ldots,v_s|a_0(k))^{1/2}}\Bigr],
\ee
where the expectation value is over measurement outcomes and choices of $M_i$ and choices of real or imaginary part.
This equals
\bea
\nonumber
&& E_{\{M_i,R_i\}}\Bigl[ \sum_{\{v_i\}} \frac{P(v_1,\ldots,v_s|a'(k))^{1/2}}{P(v_1,\ldots,v_s|a_0(k))^{1/2}} P(v_1\ldots,v_s|a_0(k))\Bigr]
\\ \nonumber
&=&  E_{\{M_i,R_i\}}\Bigl[\sum_{\{v_i\}} P(v_1,\ldots,v_s|a'(k))^{1/2} P(v_1,\ldots,v_s|a_0(k))^{1/2}\Bigr],
\eea
where the sum is over all possible sequences $v_1,...,v_s$ of measurement outcomes and the expectation value is now over all choices of $\theta_i$ and of real or imaginary part ($R_i=0,1$ is used to denote a measurement of real or imaginary part).
This equals
\bea
\label{general}
\Bigl\{ \frac{1}{t} \sum_{M=0}^{t-1} \Bigl(\frac{P(v|{\rm Re}(f_0(M)))^{1/2}  P(v|{\rm Re}(f'(M)))^{1/2}}{2} \nonumber \\+
\frac{P(v|{\rm Im}(f_0(M)))^{1/2}  P(v|{\rm Im}(f'(M)))^{1/2}}{2} \Bigr)\Bigr\}^s.
\eea

Below, we will use the assumptions on $C(x)$ to show that the term in parenthesis in Eq.~(\ref{general}) is bounded by some constant $c<1$ for all $t$.  Using this bound, the expectation value (\ref{consexgen}) is bounded by $c^s$.
Thus, for given $a'(k)$, the probability that $P(v_1,\ldots,v_s|a'(k)) \geq P(v_1,\ldots,v_s|a_0(k))$ is bounded by $c^{s}$.
Thus, the probability that there is an $a'(k)$ such that $P(v_1,\ldots,v_s|a'(k)) \geq P(v_1,\ldots,v_s|a_0(k))$ is bounded by $N_{choices} c^{s}$, as claimed.

We now bound the term in parenthesis in Eq.~(\ref{general}).
Consider $\frac{1}{t}\sum_{M}|f'(M)-f_0(M)|^2$.  This is greater than for $d_{min}^2$.  Also, for every $M$, $|f'(M)-f_0(M)|^2\leq 4A_{max}^2$.
So, for randomly chosen $M$, the probability that $|f'(M)-f_0(M)|^2$ is greater than or equal to $d_{min}^2/2$ is at least
$d_{min}^2/8A_{max}^2$.
So, the probability that if we randomly choose $M$ and randomly choose whether to measure real or imaginary part, that the corresponding part (i.e., either real or imaginary) of $f'(M)-f_0(M)$
is greater than $d_{min}/2$ in absolute value is at least $d_{min}^2/16A_{max}^2$.
Hence, by the assumption (\ref{c0assumption}) on $C(x)$, we have that
the term in parenthesis in Eq.~(\ref{general}) is bounded by
\bea
c &\leq &1-c_0 \Bigl(\frac{d_{min}}{2} \Bigr)^2 \frac{d_{min}^2}{16 A_{max}^2}.
\eea
\end{proof}
\end{thm}

\section{Kitaev's Phase Estimation Algorithm}
\label{sec:Kitaev}
Recall from Section \ref{sec:CFT} that if we apply a large number of measurements using two different values of $M$ at both $\theta=0$ and $\theta=\pi/2$, we can accurately estimate $\cos(2 \pi M \cdot\varphi_k)$ and $\sin(2 \pi M \cdot\varphi_k)$, and therefore determine $\varphi_k$.
In this section, we review Kitaev's phase estimation algorithm to determine $\varphi_k$ with exponential precision \cite{Kitaev1996} (for complete details, we refer the reader to Sec.~$13.5$ in Ref.~\onlinecite{Kitaev2002}).
This algorithm relies on obtaining accurate measurements at multiples of $\varphi_k$.  We begin by reviewing how to accurately measure a given multiple of $\varphi_k$ with constant precision, building up to estimating the phase with exponential precision.  We also simulate the algorithm to determine how many measurements are required in practice.
%using measurements at multiples $M_i$.

\subsection{Estimating $\varphi_k$ with Constant Precision}
\label{sec:KitaevMeas}
Recall that $\varphi_k = \frac{k}{t} \mod 1$, where $\varphi_k\in \mathbb{R}/\mathbb{Z}$ and $0\leq k < t <2^m$.
Let $\theta_i = \{0,\pi/2\}$ at random.
%We can think of $\theta_i$ as randomly taking the value $\pi/2$, in which case we have,
Using the measurement operator given in Section \ref{sec:KitaevMeas} and Eqs.~(\ref{cmp1},\ref{cmp2}), the conditional probability when measuring multiple $M=1$ is given by:
\be
P(0|k) = \frac{1 + \cos(2\pi\cdot \varphi_k + \theta_i)}{2}
%\nonumber
%&=& \frac{1 - \sin(2\pi\cdot\varphi_k)}{2}.
\ee
%To distinguish between $\varphi_k=\varphi$ and $\varphi_k=-\varphi$, rather than measuring $\cos$ and $\sin$ independently, we only need to randomly %assign $\theta_i = \{0,\pi/2\}$.
%%Recall that in the original Kitaev algorithm, $\theta_i = 0$ for the $\cos$ case and $\theta_i = \pi/2$ for the $\sin$ case.

We now solve for the conditional probability $P(0|k)$:
\begin{eqnarray}
&& 2P(0|k) - 1
\\ \nonumber &=& \cos(2\pi\cdot\varphi_k + \theta_i) \\ \nonumber
&=& \cos(2\pi\cdot\varphi_k) \cos\theta_i - \sin(2\pi\cdot\varphi_k) \sin\theta_i.
\end{eqnarray}

We make $s$ measurements, choosing $\theta_i \in \{0,\pi/2\}$ randomly,  to obtain approximations $P^*_{\cos}$ and $P^*_{\sin}$ close to $ \cos(2\pi\cdot\varphi_k)$ and $\sin(2\pi\cdot\varphi_k)$, respectively.
Let there be $N_c$ measurements with $\theta_i=0$.  Let $N_{c}(0)$ denote the number of these measurements having outcome $0$ and let $N_{c}(1)$ denote the number having outcome $1$.  Then, let
\be
P^*_{\cos}=\frac{N_c(0)-N_c(1)}{N_c}.
\ee
If there are $N_{s}$ measurements with $\theta_i=\pi/2$, with $N_{s}(0)$ of them having outcome $0$ and $N_{s}(1)$ having outcome $1$, then let
\be
P^*_{\sin}=\frac{N_s(1)-N_{s}(0)}{N_s}.
\ee

Given $P^*_{\cos},P^*_{\sin}$, our best estimate of $\varphi_k$ is obtained by taking an arctangent of $P^*_{\sin}/P^*_{\cos}$, choosing the appropriate quadrant.

Equivalently, we can determine multiples $M_i$ of $\varphi_k$ in the same manner by measuring and obtaining the probability
\begin{eqnarray}
P(0|k) &=& \frac{1 + \cos(2\pi M_i\cdot\varphi_k + \theta_i)}{2},
\end{eqnarray}
and computing similar estimates $P^*$ and again taking an arctangent.

In practice, how many measurements $s$ are needed to accurately determine $M_i\cdot\varphi_k$?
This is analyzed in the next two sections.

\subsection{Estimating $\varphi_k$ with Exponential Precision}
\label{KitaevAlgorithm}
To efficiently achieve exponential precision in our estimate of $\varphi_k$, we measure multiples $M_i$ of $\varphi_k$.
Then we use the measurement results in a classical inference technique to enhance the precision of the estimate.
We begin by measuring multiple $M_0 = 2^{m-1}$, then $M_1 = 2^{m-2}$,
%$2^{2n-1}\cdot\varphi_k$,
increasing the precision as we move to $M_{m-1} = 2^0$.  Each measurement gives us an estimate of $M_i \varphi_k \mo$.
%$2^0\cdot\varphi_k$.

To achieve the desired precision and probability of error, we measure each multiple $s$ times, where
in this section, $s$ refers to the number of measurements {\it per} multiple for both cosine and sine, so that the total number of measurements required is $2ms$.
The estimate of $2^{j-1}\cdot\varphi_k$, using methods of Sec.~\ref{sec:KitaevMeas}, is denoted as $\rho_j$.

We introduce binary fraction notation, where $\overline{.\alpha_1 \ldots \alpha_j} = \sum_{p=1}^j 2^{-p}\,\alpha_p,\ \alpha_p\in\{0,1\}$.
%Let $m=2n$.
%KMS: if remove 2n part, then this becomes m+1
The output of the algorithm is $\alpha = \overline{.\alpha_1\ldots\alpha_{m+2}}$, which is an exponentially precise estimate of $\varphi_k$:
\be
|\alpha - \varphi_k| < \frac{1}{2^{m+2}}.
\ee
%Estimating $2^j\cdot\varphi_k$ and the inference procedure
Kitaev's phase estimation algorithm \cite{Kitaev1996} is given in Algorithm \ref{alg:Kitaev}.
%\begin{enumerate}
%\item {\tt set} $j=m$
%\begin{enumerate}
%\item Estimate $2^{j-1} \cdot\varphi_k$ (see Sec.~\ref{sec:KitaevMeas}).
%\item Set $\beta_j$ to the octant value $\{\frac{0}{8},\frac{1}{8},...,\frac{7}{8}\}$ which is closest to the estimate of $2^{j-1} \cdot\varphi_k$.
%\item Set the three bits $\alpha_{m},\alpha_{m+1},\alpha_{m+2}$ to the bits in the binary fraction expansion of $\beta_j$.
%\end{enumerate}
%\item {\tt for} $j=m-1,\ldots,1$ {\tt do}
%\begin{enumerate}
%\item Estimate $2^{j-1} \cdot\varphi_k$ (see Sec.~\ref{sec:KitaevMeas}).
%\item Infer bit $\alpha^j$:
%\begin{equation}
%\alpha_j = \left\{
%\begin{array}{ll}
%0 & \mbox{if $|\overline{.0\alpha_{j+1}\alpha_{j+2}} - 2^{j-1}\cdot\varphi_k|_{\mod 1} < 1/4$.} \\
%1 & \mbox{if $|\overline{.1\alpha_{j+1}\alpha_{j+2}} - 2^{j-1}\cdot\varphi_k|_{\mod 1} < 1/4$.} \\
%\end{array}
%\right. \nonumber \\
%\end{equation}
%\end{enumerate}
%\item {\tt end for}
%\item {\tt return} $\alpha$, our estimate of the phase.
%\end{enumerate}

\begin{algorithm}[H]
\caption{Kitaev's Phase Estimation \cite{Kitaev1996}}
\label{alg:Kitaev}
\algsetup{indent=2em}
\begin{algorithmic}[1]
%\REQUIRE{Measurements }
%\ENSURE{$\alpha$, the estimate of the phase.}
%\STATE{Set $j=m$;}
%\STATE{Estimate $2^{m-1} \cdot\varphi_k$ (see Sec.~\ref{sec:KitaevMeas});}
%\STATE{Set $\beta_j$ to the octant value $\{\frac{0}{8},\frac{1}{8},...,\frac{7}{8}\}$ which is closest to the estimate of $2^{j-1} \cdot\varphi_k$;}
%\STATE{Set $\beta_m$ to the octant value $\{\frac{0}{8},\frac{1}{8},...,\frac{7}{8}\}$ closest to $\rho_{m-1}$;
% the estimate of $2^{j-1} \cdot\varphi_k$;}
\FOR{$j=m-1$ to $1$}
\STATE{Set $\rho_j$ to the estimate of $2^{j-1} \cdot\varphi_k$  using $O(s)$ measurements per $j$.}
\ENDFOR
\STATE{Set $\overline{.\alpha_{m}\alpha_{m+1}\alpha_{m+2}}=\beta_m$, where $\beta_m$ is the octant value $\{\frac{0}{8},\frac{1}{8},...,\frac{7}{8}\}$ closest to $\rho_{m}$.}
\FOR{$j=m-1$ to $1$}
%\STATE{Estimate $2^{j-1} \cdot\varphi_k$ (see Sec.~\ref{sec:KitaevMeas});}
\STATE{Infer $\alpha_j$:
\begin{equation*}
\alpha_j = \left\{
\begin{array}{ll}
%0 & \mbox{if $|\overline{.0\alpha_{j+1}\alpha_{j+2}} - 2^{j-1}\cdot\varphi_k|_{\mod 1} < 1/4$.} \\
%1 & \mbox{if $|\overline{.1\alpha_{j+1}\alpha_{j+2}} - 2^{j-1}\cdot\varphi_k|_{\mod 1} < 1/4$.} \\
0 & \mbox{if $|\overline{.0\alpha_{j+1}\alpha_{j+2}} - \rho_{j}|_{\mod 1} < 1/4$.} \\
1 & \mbox{if $|\overline{.1\alpha_{j+1}\alpha_{j+2}} - \rho_{j}|_{\mod 1} < 1/4$.} \\
\end{array}
\right.
\end{equation*}
}
\label{line:infer}
\ENDFOR
\RETURN{$\alpha$, the estimate of the phase.}
\end{algorithmic}
\end{algorithm}

Note that in Algorithm \ref{alg:Kitaev}, we modify the inference step in line \ref{line:infer} to use $\rho_{j}$, as opposed to using $\beta_j$ as done in Ref.~\onlinecite{Kitaev2002}.
%, where $\beta_j$ is the closest octant on the unit circle to $2^{j-1}\cdot\varphi_k$.
%Upon completion, our estimate $\alpha$ should be exponentially precise: $|\alpha - \varphi_k|< 2^{-(2n+2)}$.

\subsection{Simulation Results}
How large does $s$ need to be to estimate $\varphi_k$ to exponential precision?
The probability that a given estimate of $2^{j-1} \varphi_k$ differs by more than a given amount from the true value is exponentially small in $s$, as shown in Ref.~\onlinecite{Kitaev2002} using a Chernoff bound.  This implies that to accurately compute the word (the entire sequence of bits $\alpha$), we need $s$ to scale logarithmically with $m$.
%Here we present numerical results to more precisely quantify this scaling.

\begin{figure*}[htb]
  \centering
  \subfloat[][1000 Bits.]{
 \includegraphics[width=3.5in]{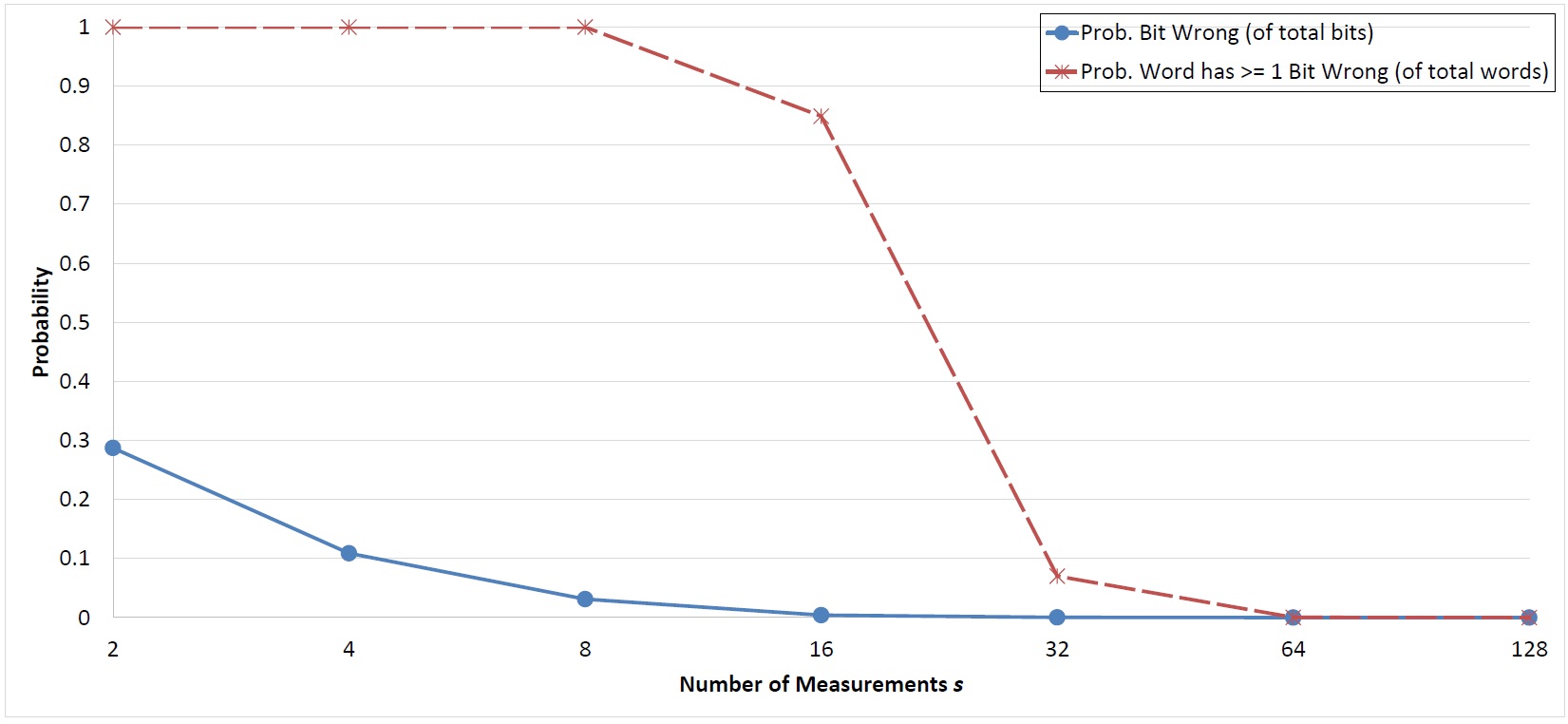}
  \label{fig:Kitaev1000}
}
  \subfloat[][10000 Bits.]{
 \includegraphics[width=3.5in]{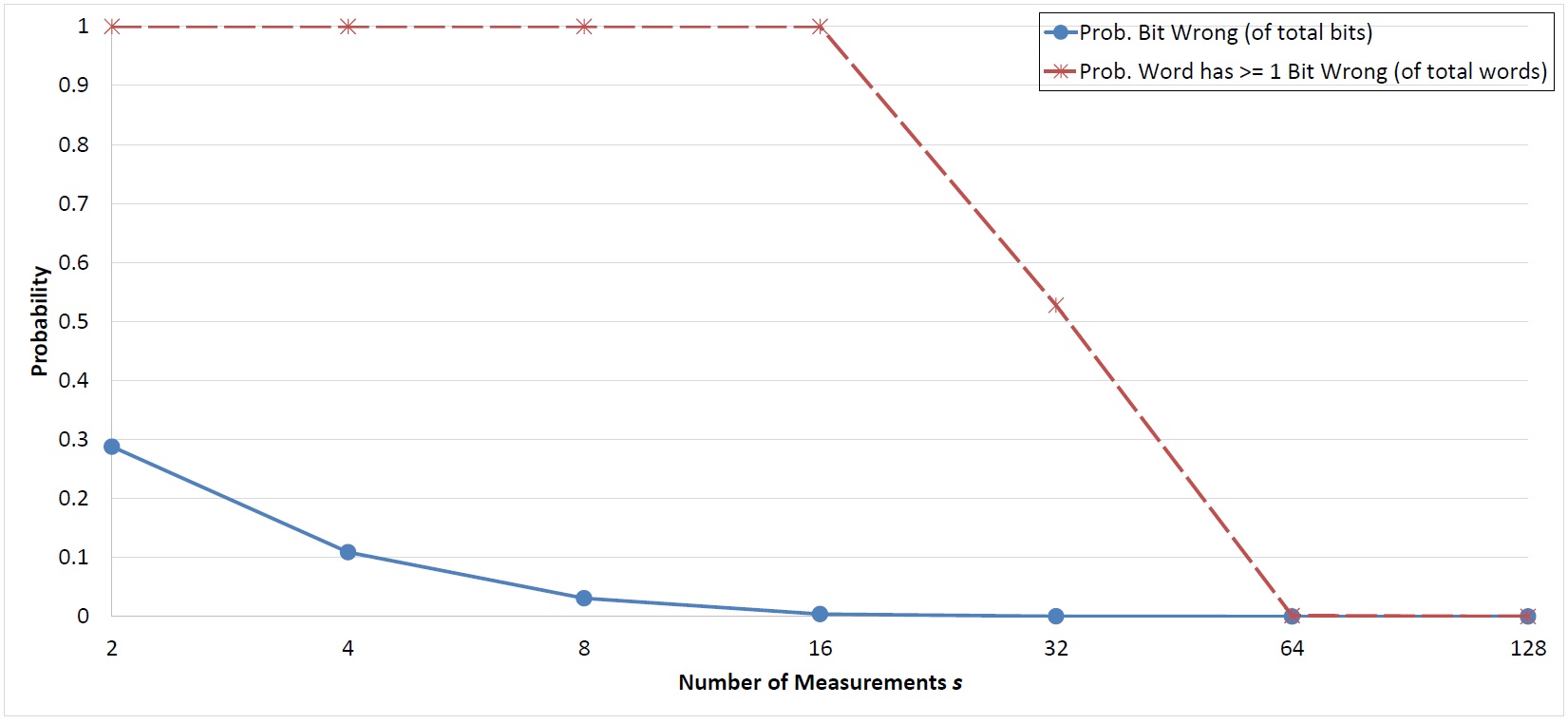}
  \label{fig:Kitaev10000}
}
  \caption{The number of measurements $s$ versus the probability of a given bit being wrong (blue) and a given word being wrong (red), meaning probability that a given word has at least one bit wrong.  (a) Simulation results for words of length 1000 bits.  (b) Simulation results for words of length 10000 bits.}
  \label{fig:Kitaev}
\end{figure*}

We ran $10000$ independent simulations of Algorithm \ref{alg:Kitaev}, for words of length $m=\{1000,10000\}$.
These word lengths are of particular interest since Shor's algorithm promises computational speed-ups over its classical counterpart for word lengths around $2048$--$4096$.
We considered performance of the algorithm as we varied the number of measurements $s$ of each multiple.
Figure \ref{fig:Kitaev} shows the numerical results.
The $x$-axis is the number of measurements $s$.
The $y$-axis is the probability, where
we plot both the probability of a given bit being wrong (blue) across all bits and simulation runs, and the probability of a given word being wrong (red) across all simulation runs (i.e., the probability that at least one bit in the word is wrong).
For both word lengths, we see that $s$ scales logarithmically in $m$, and does not exceed $64$.
The number of measurements required thus scales as $O(m\log(m))$.
The corresponding classical post-processing circuit scales as $O(m)$ size and $O(\log(m))$ depth.

%\begin{figure}[htb]
%  \centering
%  \includegraphics[width=3.5in]{Kitaev_1000Bit.pdf}
%  \caption{Scaling of $s$ relative to probability of wrong bits.}
%  \label{fig:Kitaev}
%\end{figure}

%Table \ref{fig:1000Table} contains the raw data for the plot.
%
%\begin{figure}[htb]
%  \centering
%  \includegraphics[width=3.5in]{Kitaev_1000BitData.pdf}
%  \caption{Scaling of $s$ relative to probability of wrong bits.}
%  \label{fig:1000Table}
%\end{figure}

%\begin{eqnarray}
%\cos \varphi &=& \frac{1}{N} \sum_{\cos meas} X_i\\
%\sin \varphi &=& \frac{1}{N} \sum_{\sin meas} X_i
%\end{eqnarray}

\section{Fast Phase Estimation}
\label{sec:FastKitaev}
In this section, we extend Kitaev's algorithm for phase estimation by considering inference across multiple bits simultaneously.
We begin by describing an algorithm that improves the number of measurements $O(m \log(m))$ in the previous section to $O(m \log(\log(m))$.  This algorithm consists of two ``rounds", where the first
round is similar to Kitaev's algorithm, but the second round infers multiple bits simultaneously.  Having described this algorithm, we then describe how to further improve it by considering more rounds, requiring $O(m \log(\log(\log(m))))$ measurements for three rounds, and so on, ultimately describing an algorithm that requires
$O(m \log^*(m))$ measurements, where $\log^*(m)$ is the iterated logarithm and is bounded for all practical purposes by $5$.
These algorithms all require only an amount of computational time for classical post-processing that is $O(m \log(m))$ as discussed at the end of the section.

The algorithms in this section can be motivated as follows: the limitation of Kitaev's algorithm is that it infers single bits at a time, and requires logarithmically many measurements per bit.  So, a natural generalization is to consider multiples $M$ that are not powers of two, so that we can infer multiple bits at a time.  The information theory method does this, by using random $M$, but requires lots of post-processing.  So, in this section we consider ``sparse" $M$, in that the $M$ will be a sum of a small number of powers of two.  There is a tradeoff, in that as the ``density" (defined to be the number of powers of two) increases, the number of measurements required is reduced, but the postprocessing becomes more complicated.  So to make the inference efficient, we use a bootstrapping procedure with multiple rounds, with the density increasing from one round to the next.  The early rounds yield only imperfect inferences, but they give enough information to simplify the inference in later rounds.

\subsection{Two Round Algorithm}
The measurements that we use in the first round of the two-round algorithm are equivalent to those of Kitaev's algorithm, except that the parameter $s$ will be chosen differently.  We set $s=s_1$ for some $s_1$ chosen later (we call this quantity $s_1$ as in the second round we will have an $s_2$, and so on).
Using a Chernoff bound estimate as in Ref.~\onlinecite{Kitaev2002}, we can bound the probability that the difference between $2^{j-1}\cdot\varphi_k$ and our best estimate of $2^{j-1}\cdot\varphi_k$ is greater than $1/16$ by  $\exp(-c s_1)$
for some constant $c>0$.
For notational simplicity, we will use one piece of notation that was used in Kitaev's original algorithm: for each $j$, we will let $\beta_j$ be the closest approximation in the set $\{\frac{0}{8},...,\frac{7}{8}\}$ to the estimate of $2^{j-1} \cdot\varphi_k$.  So, the probability of an error larger than $1/8$ in $\beta_j$ is bounded by $\exp(-c s_1)$.

In the original Kitaev algorithm, we then combine these $\beta_j$ to estimate $\varphi_k$.
Instead, our goal in the first round of the two-round algorithm is to give, for almost every $j$, a quantity called $\rho_j$ that will be an estimate of $2^{j-1}\cdot \varphi_k$ to a precision $\delta_1$, where the subscript $1$ is to indicate that this is the precision on the first round.  This quantity $\delta_1$ will be much larger than the final precision $\delta$ of our two-round algorithm, but will be much smaller than $1$.
We say ``almost every" $j$ because, as we will see, we will only be able to give this precise estimate $\rho_j$ for $0 \leq j<m-\log(1/\delta_1)$; however, since $\log(\delta_1)$ will be much smaller than $m$, this will indeed be
most of the $j$.
To compute $\rho_j$,
 we use $\beta_{j+l}$ for $l=0,...,\log(1/\delta_1)$ in a Kitaev-style inference procedure to compute $\log(1/\delta_1)+2$ bits in the binary expansion of $\rho_j$.  That is, we obtain the three lowest order bits in the binary expansion from $\beta_{j+\log(1/\delta_1)}$.  We then sharpen the estimate, obtaining the $l^{\textrm{th}}$ bit in the binary expansion from $\beta_{j+l-1}$ and from the $l+1^{\textrm{th}}$ and $l+2^{\textrm{th}}$ bits, proceeding iteratively.
We can bound the probability of error in $\rho_j$ by
\be
\label{deltaEqFast}
{\rm Pr}\Bigl[ \Bigl|\rho_j-2^{j-1}\cdot \varphi_k\Bigr|\smo \geq \delta_1 \Bigr] \leq \log(1/\delta_1) \exp(-c s_1).
\ee
The factor of $\log(\delta_1)$ occurs because to obtain an error less than $\delta_1$ requires $\log(1/\delta_1)$ bits of precision.
This estimate of the probability of error is essentially the same as the estimate of the probability of having an error in Kitaev's original algorithm, except that instead of having $m$ bits in the expansion, we have $\log(1/\delta_1)$ bits.  The event of having large error for some given $j$ is uncorrelated with the event of having large error for bits $j'$ if $|j'-j|$ is large enough compared to $\log(1/\delta_1)$.  This will play an important role in analyzing the algorithm later, allowing us to neglect certain correlations (we will explain this below, although we will not give a mathematical proof of this).
The fact that we have only obtained the accurate estimate of $\rho_j$ for $j\leq m-\log(1/\delta_1)$ will not pose a difficulty in what follows; this will be only a minor technical detail.  For one thing, most ``sets of measurements" (as defined below) do not ``contain" (also defined below) the $j$ for which we do not have an accurate estimate.  Alternatively, we can simply on the first round infer all $\rho_j$ for $j \leq m$ accurately by running the first round on $m+\log(1/\delta_1)$ bits.

The second round uses $s_2 m$ ``sets" of measurements, for some parameter $s_2$ chosen later, where each set of measurements will consist of repeating the same measurement a total of $C$ times, for some constant $C$.  We also introduce a parameter $S$, called the ``density" in this round.
For the $i^{\textrm{th}}$ set of measurements, we pick $S$ different random values of $j$ in the range $1 \leq j \leq m$, calling these values $j^i_1,...,j^i_S$.  We will require that these values $j^i_1,...,j^i_S$ all be distinct from each other in a given measurement (if any two are equal, we simply generate another $S$-tuple of values; we will have $S << \sqrt{m}$ so a random tuple will have distinct entries with probability close to $1$).
Then we estimate
\be
\Bigl( 2^{j^i_1-1} + 2^{j^i_2-1}  + ... + 2^{j^i_S-1} \Bigr) \varphi_k,
\ee
calling this estimate $\sigma_i$.
We do this estimate using $C$ applications of the basic measurement operation, with $M_i= 2^{j^i_1-1} + ... + 2^{j^i_S-1}$ for each measurement and $\theta_i$ being chosen randomly in $\{0,\pi/2\}$.  The constant $C$ will be chosen so that
\be
\label{PFail2ndRound}
{\bf Pr}\Bigl[\Bigl|M_i \cdot\varphi_k - \sigma_i\Bigr|\smo > 1/32 \Bigr] \leq \frac{1}{8}.
\ee
The constant $C$ is of order unity and does not depend upon $m$.

This completes the description of the measurements in the two-round algorithm.  We now describe the classical post-processing phase.  We will explain below how to estimating a quantity $\beta'_j$ for each $j$.  This quantity will be an approximation to $2^{j-1} \varphi_k$, chosen from the set $\{\frac{0}{8},...,\frac{7}{8}\}$.  The goal of the algorithm is to obtain an estimate such that
for all $j$ we have
\be
{\bf Pr}\Bigl[\Bigl|2^{j-1} \cdot\varphi_k - \beta'_j \Bigr|\smo > 1/16 \Bigr]
\leq \frac{\epsilon}{m},
\ee
for some constant $\epsilon$.  Thus, by a union bound, the probability of an error greater than $1/8$ in any of the $\beta'_j$ will be bounded by $\epsilon$.  We then use the $\beta'_j$ to determine the $\alpha_j$ using a procedure similar to Kitaev's algorithm.  This procedure is given in Algorithm \ref{alg:fast}, steps $13-17$.
%\begin{enumerate}
%\item Set the three bits $\alpha_{m},\alpha_{m+1},\alpha_{m+2}$ to the bits in the binary fraction expansion of $\beta'_j$.
%\item {\tt for} $j=m-1,\ldots,1$ {\tt do}
%\begin{enumerate}
%\item Infer $\alpha_j$:
%\begin{equation}
%\alpha_j = \left\{
%\begin{array}{ll}
%0 & \mbox{if $|\overline{.0\alpha_{j+1}\alpha_{j+2}} - \beta'_k|_{\mod 1} < 1/4$.} \\
%1 & \mbox{if $|\overline{.1\alpha_{j+1}\alpha_{j+2}} - \beta'_k|_{\mod 1} < 1/4$.} \\
%\end{array}
%\right. \nonumber \\
%\end{equation}
%\end{enumerate}
%\item {\tt end for}
%\item {\tt return} $\alpha$, our estimate of the phase.
%\end{enumerate}
%
\begin{algorithm}[H]
\caption{Fast Phase Estimation}
\label{alg:fast}
\algsetup{indent=2em}
\begin{algorithmic}[1]
\STATE{First Round:}
\FOR{$j=m-1$ to $1$}
\STATE{Estimate $2^{j-1} \cdot\varphi_k$  using $O(1)$ measurements per $j$.}
\ENDFOR
\STATE{Later Rounds:}
\FOR{$r=2$ to Number of Rounds}
\STATE{Set density, $S$, and number of measurements per bit, $s_r$, for given round.}
\FOR{$i=1$ to $s_r m$}
\STATE{Set $M_i$ to a sum of $S$ different powers of two, choosing these powers of two at random or with a pseudo-random distribution.  Perform $O(1)$ measurements with given $M_i$ and random or pseudo-random $\theta$.}
\ENDFOR
\ENDFOR
\STATE{Perform multi-bit inference to determine estimate of $\beta'_j=2^{j-1}\cdot \varphi_k$ for all $j$.  Use estimates from previous round to give starting point for inference in next round.  See text for details.}
\STATE{Set $\overline{.\alpha_{m}\alpha_{m+1}\alpha_{m+2}}=\beta'_m$;}
\FOR{$j=m-1$ to $1$}
%\STATE{Estimate $2^{j-1} \cdot\varphi_k$ (see Sec.~\ref{sec:KitaevMeas});}
\STATE{Infer $\alpha_j$:
\begin{equation*}
\alpha_j = \left\{
\begin{array}{ll}
0 & \mbox{if $|\overline{.0\alpha_{j+1}\alpha_{j+2}} - \beta'_j|_{\mod 1} < 1/4$.} \\
1 & \mbox{if $|\overline{.1\alpha_{j+1}\alpha_{j+2}} - \beta'_j|_{\mod 1} < 1/4$.} \\
\end{array}
\right.
\end{equation*}
}
\label{line:inferfast}
\ENDFOR
\RETURN{$\alpha$, the estimate of the phase.}
\end{algorithmic}
\end{algorithm}
If the error is bounded by $1/8$ for all $\beta'_j$, then the estimate of the phase will be accurate to $2^{-(2n+2)}$.

To estimate $2^{j-1}\cdot\varphi_k$ for a given $j$, consider all sets of measurements such that one that one of the random values of $j_a$ was equal to $j$; we say that such a set of measurements ``contains $j$".
On average, there will be $s_2 S$ such sets of measurements.  Let us first proceed by assuming that there are indeed exactly $s_2 S$ sets of measurements and then later deal with the fluctuations in the number of sets of measurements.
On the $i^{\textrm{th}}$ set of measurements, we obtain some estimate of $\sigma_i$.  Suppose this set contains $j$.
 Without loss of generality, let us suppose that $j_1=j$.
Then, given only $\sigma_i$ and $\rho_{j_2},...,\rho_{j_S}$, our best estimate of $2^{j-1} \cdot\varphi_k$ is:
\be
\label{BestEst}
\sigma_i-\rho_{j_2}-\rho_{j_3}-...-\rho_{j_S}
\ee

We now bound the probability that the estimate is off by more than $1/16$.  We do this by bounding the probability that our value of $\sigma_i$ differs by more than $1/32$ from the correct value using Eq.~(\ref{PFail2ndRound}) and also bounding the probability that our estimate of $\sum_{l=2}^S \rho_{j_l}$ differs by more than $1/32$ from the correct value.  To bound that probability,
we have
\be
\label{PFail}
{\rm Pr}\Bigl[ \Bigl|\Bigl( \sum_{l=2}^S  \rho_{j_l} - 2^{j_l-1} \cdot\varphi \Bigr) \Bigr|\smo \geq \frac{1}{32} \Bigr] \leq S \log(32 S) \exp(-c s_1),
\ee
where we have taken $\delta_1=1/32 S$ in Eq.~(\ref{deltaEqFast}) so that if each quantity $\rho_{j_l} - 2^{j_l-1} \cdot\varphi $ is accurate to within $\delta_1$ then the sum is accurate to within $1/32$.
We then use a union bound: if the probability that any given measurement is inaccurate is bounded by $\log(1/\delta_1) \exp(-c s_1)$, then the probability that at least one measurement is inaccurate is bounded by $S$ times that quantity.

We choose $s_1 \sim \log(\log(m))$ and $S \sim \log(m)$ so that the right-hand side of Eq.~(\ref{PFail}) is bounded by $1/32$.
Then, using Eqs.~(\ref{PFail2ndRound},\ref{PFail}), the probability that the quantity in Eq.~(\ref{BestEst}) differs by at least $1/16$ from $2^{j-1} \cdot\varphi_k$ is bounded by $1/4$.
We get roughly $s_2 S$ different estimates of $2^{j-1} \cdot\varphi_k$, one for each set of measurements involving the given $j$.  Let us assume the independence of certain events between different sets of measurements, namely the event that the quantity in Eq.~(\ref{BestEst}) differs by more than $1/32$ from $2^{j-1} \cdot\varphi_k$ (we discuss this further below).
Then, we can combine these measurements to obtain an estimate of $\beta'_j$ by picking the value of $\beta'_j$ which is most frequently within $1/16$ of
$\sum_{l=2}^S \Bigl( \rho_{j_l} - 2^{j_l-1} \cdot\varphi \Bigr)$; i.e., it is within $1/16$ of that value for the greatest number of sets of measurements containing $j$.

The probability of error in $\beta'_j$ by more than $1/16$ is then bounded by $\exp(-c' s_2 S)$ for some constant $c'>0$.  Picking $s_2 \sim 1$, we find that the probability of error is $1/{\rm poly}(m)$ for any desired polynomial, with the power depending upon the ratio between $S/\log(m)$, so we can ensure that this probability is small compared to $\epsilon/m$.
The number of measurements required by this procedure is $O(m \log(\log(m)))$.

We now discuss several issues of correlations and fluctuations that were left open in the above analysis.  First, consider the fluctuation in the number of sets of measurements
that contain $j$, for any given $j$.  On average this quantity is $s_2 S$, but there may be some fluctuations.  However, the probability that there are fewer than $s_2 S/2$ different such sets of measurements is exponentially small in $s_2 S$, and hence for the given choice of $s_2 S$, the quantity is bounded by $1/{\rm poly}(m)$ and so can be made negligible (in fact, this probability, being exponentially small in $s_2 S$, has a similar scaling as the probability that we incorrectly infer a given $2^{j-1} \cdot\varphi_k$ given $s_2 S$ sets of that contain $j$, as that probability is also exponentially small in $s_2 S$).  So, with high probability all $j$ are contained in at least $s_2 S/2$ measurements, and so we can double $S$ and apply the analysis above.

It is possible that a better way to deal with fluctuations in the number of measurements is to change the distribution of choices of $j_a^i$, and anti-correlate the choices in different sets of measurements to reduce the fluctuations in the number of sets of measurements containing a given $j$.  This will at best lead to a constant factor improvement.

Another kind of correlation that we must deal with is correlation between the events that the quantity in Eq.~(\ref{BestEst}) differs by more than $1/16$ from $2^{j-1} \cdot\varphi_k$.  For a given $j$, let us assume for a given set of measurements we have $j^i_1=j$.  Let us refer to $j^i_2,...,j^i_S$ as the ``partners" of $j$.  For a given $j$, the different sets of measurements involving that $j$ will typically have wildly different partners of $j$; that is, for two different sets of measurements, $m,n$, we will typically have $|j^m_a-j^n_b| \gtrsim m/S^2 >> \log(S)$ for $a,b \neq 1$.  So, for most sets of measurements, these will be independent.  Similarly, in a given measurement we will typically have $|j^m_a-j^m_b| >> \log(S)$ so we can ignore correlations between errors in different $\rho_{j_a}$.

Of course, the above is not a rigorous proof, but we expect that such a proof can be provided without any significant difficulty.  Note that if for a given $j$ we have a large number of (roughly) independent sets of measurements containing that $j$, then adding a small number of correlated sets of measurements will not prevent the inference from working.

\subsection{Multiple Round Algorithm}
We can improve this procedure by increasing the number of rounds.  In the first and second rounds we proceed as before, though the constants $s_1,s_2,S$ will be changed.  Let us write $S=S_2$ for the second round.  The third round of the procedure is the same as the second round, except that we do $s_3$ sets of measurements, and in each measurement we pick $S_3$ different random values of $j$.  On the third round, as in the second, we repeat each set of measurements $C$ times; it is only the first round where the quantity $C$ does not appear, for the reason that in that round, each measurement is already being repeated $s_1$ times.  We can increase the number of rounds indefinitely.  In each round, we can exponentially increase the density compared to the previous round, while keeping all constants $s_a$ of order unity.  The number of measurements required is then proportional to the number of rounds.  Since $S$ increases exponentially in each round and we need $S \sim \log(m)$ in the last round, the number of rounds required is $\sim \log^*(m)$ and the total number of measurements is $\sim m \log^*(m)$.

\subsection{Classical Post-processing Time Required}
The simplest implementation of the algorithm above requires a time $O(m \log^2(m))$.  We discuss this first and then discuss how to improve to $O(m \log(m))$.  Each bit is contained in $\sim\log(m)$ sets of measurements (indeed, the fact that it is contained in this many sets of measurements is the whole point of the algorithm).  To compute the quantity in Eq.~(\ref{BestEst}), the sum on the right-hand side require summing over $S$ different quantities, and for $S \sim \log(m)$, this means that it takes time $\sim \log(m)$ to do the computation for each bit for each set of measurements containing that bit.  So, with $m$ bits, each contained in $~\log(m)$ sets of measurements, the time is $O(m \log^2(m))$.

However, we can slightly improve this by noting that Eq.~(\ref{BestEst}) can be written as
\be
\label{BestEst2}
\sigma_i-\Bigl(\rho_{j_1}+\rho_{j_2}+...+\rho_{j_S} \Bigr) + \rho_{j_1}.
\ee
The quantity in parentheses can be computed once for each set of measurements, and re-used in inferring each of the $\rho_{j_i}$ for $i \in \{1,...,S\}$, and then it only requires $O(1)$ time to do the arithmetic for each of these $i$.  This improves the total time to $O(m \log(m))$.
%Additional notes:
%\begin{eqnarray}
%\cos(2\pi.\alpha_j \alpha_{j+1} \alpha_{j+2})\\
%= \cos(2\pi.\alpha_j)\cos(\alpha_{j+1}\alpha_{j+2})\\
%\sin(2\pi.\alpha_j \alpha_{j+1} \alpha_{j+2})\\
%= \sin(2\pi.\alpha_j)\sin(\alpha_{j+1}\alpha_{j+2})\\
%\end{eqnarray}

%Steps for the code:
%\begin{enumerate}
%\item Create identical program to Kitaev version.  Reproduce results, using new function calls and updates.
%\item $\alpha_{j+1}\alpha_{j+2} \rightarrow \theta.b []_{0/1} []_{cos}$
%\item Maintain a running sum.
%\item Infer within the loop.
%\end{enumerate}

%\section{Experiments and Results}

%\section{Conclusion and Future Work}

% The \nocite command causes all entries in a bibliography to be printed out
% whether or not they are actually referenced in the text. This is appropriate
% for the sample file to show the different styles of references, but authors
% most likely will not want to use it.
%\nocite{*}

\section{Analysis of Quantum Circuit Depth and Width}
\label{sec:circuits}
The fast phase estimation algorithm offers an asymptotic improvement in the number of measurements required to estimate the phase with exponential precision.
How does the corresponding quantum circuit scale, in terms of depth, width and size?
We define the {\em depth} of a quantum circuit as the number of timesteps, where gates on disjoint qubits can occur in parallel in a given timestep.
Here we assume that a given $n$-qubit gate takes one timestep.
The {\em width} of a quantum circuit is the number of qubits.
The {\em size} of a quantum circuit is the total number of non-identity quantum gates.
%The {\em area} $A$ of a quantum circuit is roughly the depth times the width.
We analyze the circuits given three different computing settings, to emphasize tradeoffs in depth and width depending on resource availability.
Table \ref{table:circuits} contains a summary of the circuit resources required for each algorithm given the setting.

\begin{table*}[tb]
\caption{Table of circuit depth, width, and size for Kitaev's quantum phase estimation and fast phase estimation.}
\begin{tabular}{|c||c|c|c||c|c|c|}\hline
\multirow{2}{*}{Type} & \multicolumn{3}{c||}{Kitaev's Phase Estimation \cite{Kitaev2002}} & \multicolumn{3}{c|}{Fast Phase Estimation} \\ \cline{2-7}
 & Depth & Width & Size & Depth & Width & Size \\ \hline
Sequential & $O(m\log(m))$ & $O(m\log(m))$ & $O(m\log(m))$ & $O(m\log^*(m))$ & $O(m\log^*(m))$ & $O(m\log^*(m))$ \\ \hline
Parallel & $O(\log(m))$ & $O(m\log(m))$ & $O(m\log(m))$ & $O(\log(m))$ & $O(m\log^*(m))$ & $O(m\log^*(m))$ \\ \hline
Cluster & $O(\log(m))$ & $O(m^2)$ & $O(m\log(m))$ & $O(\log^*(m))$ & $O(m^2)$ & $O(m\log^*(m))$ \\ \hline
\end{tabular}
\label{table:circuits}
\end{table*}

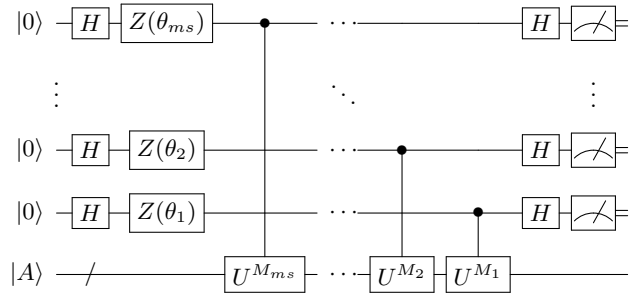
\begin{figure}[tb]
\centering
\[
\Qcircuit @C=.5em @R=1em {
\lstick{\ket{0}}& \gate{H} & \gate{Z(\theta_{ms})}  &\ctrl{4} & \qw & \push{\cdots}  & \qw & \qw & \gate{H}    & \meter    & \cw \\
\push{\vdots} & & & & & \push{\ddots} & & & & \push{\vdots}&\\
\lstick{\ket{0}} & \gate{H}   & \gate{Z(\theta_2)}  & \qw &\qw & \push{\cdots} & \ctrl{2} &\qw  & \gate{H}    & \meter    & \cw \\
\lstick{\ket{0}} & \gate{H}   & \gate{Z(\theta_1)}    & \qw & \qw & \push{\cdots} & \qw & \ctrl{1}   & \gate{H}    & \meter    & \cw \\
\lstick{\ket{A}} & {/}\qw  &\qw    & \gate{U^{M_{ms}}} & \qw & \push{\cdots} & \gate{U^{M_2}} & \gate{U^{M_1}} &    \qw      &    \qw    & \qw }
\]
\caption{Quantum circuit for sequential phase estimation.}
\label{fig:seq}
\end{figure}

First, consider the setting where each measurement operator is performed sequentially.
That is, the circuit is given by the sequence of gates (shown in Fig.~\ref{fig:seq})
\bea
\label{circuitSeq}
H^{\otimes ms }\, &&Z(\theta_{M_1})\Lambda^1(U^{M_1})[q_1,A]\, Z(\theta_{M_2})\Lambda^1(U^{M_2})[q_2,A] \ldots \nonumber\\ &&Z(\theta_{M_{ms}})\Lambda^1(U^{M_{ms}})[q_{ms},A]\, H^{\otimes ms},
%H^{\otimes ms }\left(\prod_{i=1}^{ms}Z(\theta_{M_i})\Lambda^1(U^{M_i})[q_i,A]\right) H^{\otimes ms},\nonumber
\eea
where $\Lambda^n(U)[q_1,q_2]$ denotes $n$-qubits in register $q_1$ controlling the application of gate $U$ to register $q_2$.
The quantum register containing the eigenvector state is denoted by $\ket{A}$ and consists say of $a$ qubits.
Each phase estimation algorithm performs $O(ms)$ measurements, resulting in a circuit of depth and size $O(ms)$.
The circuit requires $O(ms)$ ancilla qubits, one per measurement, plus $a$ additional qubits.
For Kitaev's phase estimation and fast phase estimation, $s$ equals $O(\log(m))$ and $O(\log^*(m))$, respectively.
Thus in the sequential setting, fast phase estimation offers an asymptotic improvement in circuit depth and size, as well as in the number of ancilla qubits.

Second, consider a more parallel setting obtained by decreasing circuit depth at the cost of increasing circuit width.
We can {\it parallelize} quantum phase estimation using techniques presented in Refs.~\onlinecite{Kitaev2002, CW00}.
The idea is to apply one multi-controlled gate instead of the sequence in Eq.~(\ref{circuitSeq}), by evolving as
\be
\ket{M}\otimes\ket{A} \rightarrow \ket{M} \otimes U^M\ket{A},
\ee
where $M$ is given by the sum of the multiples:
\be
M = \sum_{j=1}^{ms} M_j q_j.
\ee
%, rather than the sequence given in (\ref{circuitSeq}).
%In our case, $q_j=1$ for all $j$.
%\be
%\Lambda^{|M|}(U^M)[,A],
%\ee
%where $|p|$ is the number of bits in $p$, and $q_1$ is a register containing $|p|$ qubits.

The sum can be computed using a quantum addition circuit based on a 3-2 quantum adder (also called a carry-save adder) \cite{Kitaev2002,Pham2012,Gossett1998},
which reduces the sum of three $m$-bit numbers to a sum of two encoded numbers in $O(1)$ depth, with width $O(m)$ and size $O(m)$.
Consider $M$ to be a sum of $s$ $m$-bit integers.
The circuit first uses a $\log(s)$-depth tree of 3-2 adders to produce two encoded numbers,
and then adds these two numbers in place using a quantum carry-lookahead adder \cite{Draper2004} with $O(m)$ ancillae, $O(m)$ size, and $O(\log(m))$ depth.
In total, the addition requires a quantum circuit of $O(ms)$ ancillae, $O(\log(s) + \log(m))$ depth, and $O(ms)$ size.
%Then a quantum circuit can be constructed to add $s$ $m$-bit integers in depth $O(\log m + \log s)$ and $O(ms)$ size and width, by first using a tree of 3-2 quantum adders, resulting in two %$m$-bit numbers.  Finally, the two $m$-bit numbers can be add using a quantum carry-lookahead adder \cite{Draper2004}.
%(for details, we refer the reader to Exercise 2.13 in Ref.~\onlinecite{Kitaev2002}).

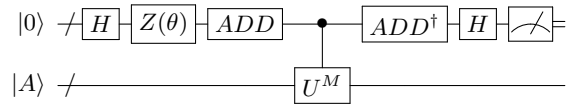
\begin{figure}[tb]
\centering
\[
\Qcircuit @C=.5em @R=1em {
%\lstick{\ket{0}} & {/} \qw       & \multigate{1}{ADD} & \qw & \qw & \qw & \multigate{1}{ADD^\dagger} & \qw \\
\lstick{\ket{0}} & {/} \qw & \gate{H} & \gate{Z(\theta)}& \gate{ADD}  &\ctrl{1}   & \gate{ADD^\dagger} & \gate{H}    & \meter    & \cw \\
\lstick{\ket{A}} & {/} \qw &\qw       & \qw               & \qw         & \gate{U^M} &    \qw            &    \qw    & \qw & \qw}
\]
\caption{Quantum circuit for parallel phase estimation.  Each wire represents a register of qubits.}
\label{fig:parallel}
\end{figure}

The circuit for performing parallel phase estimation is shown in Figure \ref{fig:parallel}.
The circuit begins with a quantum register containing qubits initialized to $\ket{0}$.  Each qubit $q_i$ undergoes a Hadamard operation, followed by a phase rotation by angle $\theta_i$ about the $z$-axis.
An addition circuit is applied to determine $\ket{M}$.
%The addition circuit can be computed with a circuit of size $O(ms)$ and depth $O(s)$.
A controlled $U^M$ operation is applied, followed by an addition circuit to uncompute $\ket{M}$.
Finally, $O(ms)$ Hadamard operations and measurements are applied, which can be done in depth $O(1)$.
The complete circuit for parallel phase estimation requires $O(ms)$ size and $O(\log(s) + \log(m))$ depth, up to the implementation of the multi-controlled $U^M$ gate.
Again, $s$ equals $O(\log(m))$ for Kitaev's phase estimation and $O(\log^*(m))$ for fast phase estimation yielding a significant reduction in circuit size and width to $O(m\log^*(m))$.

Third, consider access to a cluster of quantum computers containing $m$ nodes.
Each node performs $s$ measurements, resulting in a depth of $O(s)$, with a size and width {\em per} node of $O(s)$ gates and $O(s + a)$ qubits, respectively.
The cumulative cost across all $m$ nodes is $O(s)$ depth, $O(ms)$ size, and $O(ms + ma)$ qubits.
Again, fast phase estimation yields asymptotic improvements in all dimensions, and results, for all practical purposes, in a constant-depth phase estimation circuit.
%Phase estimation requires $s=O(\log(m))$ measurements, while fast phase estimation requires $s=O(\log^*(m))$ measurements.
%
%KMS: double check scaling of qubits in the 3-2 adder.
%Interestingly, although this system requires that each node in the cluster prepare the state $|A\rangle$, it does not require the additional ancilla needed for addition, resulting in better scaling than the parallel methods.
%Another
One potential advantage of the cluster model is that errors do not accumulate on the eigenvector state $|A\rangle$, since subsets of measurements are done on separate nodes.
This could be advantageous when designing a fault-tolerant phase estimation algorithm.

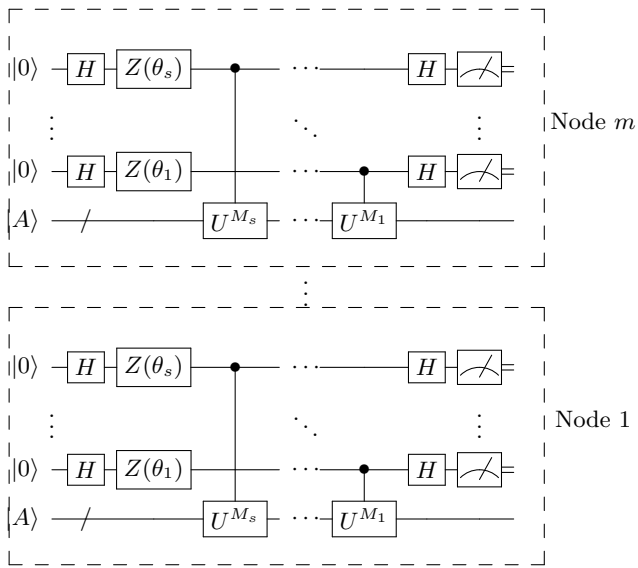
\begin{figure}[tb]
\centering
\[
\Qcircuit @C=.5em @R=.5em {
\lstick{\ket{0}} & \gate{H} & \gate{Z(\theta_{s})}   & \ctrl{3} &\qw& \push{\cdots}  & \qw                    & \gate{H} & \meter & \cw &\\
\push{\vdots}    &          &                        &          && \push{\ddots}  &                        &          & \push{\vdots}& &&&\push{\mbox{Node $m$}}\\
\lstick{\ket{0}} & \gate{H} & \gate{Z(\theta_1)}     & \qw      &\qw& \push{\cdots}  & \ctrl{1}               & \gate{H} & \meter & \cw \\
\lstick{\ket{A}} & {/}\qw   &\qw                     & \gate{U^{M_{s}}}&\qw& \push{\cdots}  & \gate{U^{M_1}}  &\qw       & \qw    & \qw \\
\\
\lstick{} & & & & & \push{\vdots} & & & & &\\
\\
\\
\lstick{\ket{0}} & \gate{H} & \gate{Z(\theta_{s})}   & \ctrl{3} &\qw& \push{\cdots}  & \qw                    & \gate{H} & \meter & \cw &\\
\push{\vdots}    &          &                        &          && \push{\ddots}  &                        &          & \push{\vdots}&&&&\push{\mbox{Node $1$}}\\
\lstick{\ket{0}} & \gate{H} & \gate{Z(\theta_1)}     & \qw      &\qw& \push{\cdots}  & \ctrl{1}               & \gate{H} & \meter & \cw \\
\lstick{\ket{A}} & {/}\qw   &\qw                     & \gate{U^{M_{s}}}&\qw& \push{\cdots}  & \gate{U^{M_1}}  &\qw       & \qw    & \qw
\gategroup{1}{1}{4}{9}{3.5em}{--}
\gategroup{9}{1}{12}{9}{3.5em}{--}
%\gategroup{6}{1}{7}{8}{1.9em}{--}
}
%
%\lstick{} & \lstick{\ket{0}} & \gate{H} &\gate{Z(\theta)}    &\ctrl{1}   & \gate{H}    & \meter    & \cw & \push{\mbox{Node $m$}}\\
%\lstick{} & \lstick{\ket{A}} &   {/}  \qw & \qw      & \gate{U^{M_m}} &    \qw      &    \qw    & \qw\\
%\lstick{} & & & & \push{\vdots} & & & \\
%\lstick{} &\lstick{\ket{0}} & \gate{H}  &\gate{Z(\theta)}   &\ctrl{1}   & \gate{H}    & \meter    & \cw & \push{\mbox{Node $2$}}\\
%\lstick{} &\lstick{\ket{A}} &   {/}  \qw & \qw      & \gate{U^{M_2}} &    \qw      &    \qw    & \qw\\
%\lstick{} &\lstick{\ket{0}} & \gate{H}  &\gate{Z(\theta)}   &\ctrl{1}   & \gate{H}    & \meter    & \cw & \push{\mbox{Node $1$}}\\
%\lstick{} &\lstick{\ket{A}} &   {/}  \qw & \qw      & \gate{U^{M_1}} &    \qw      &    \qw    & \qw
%\gategroup{1}{1}{2}{8}{1.9em}{--}
%\gategroup{4}{1}{5}{8}{1.9em}{--}
%\gategroup{6}{1}{7}{8}{1.9em}{--}
\]
\caption{Quantum circuit for parallel phase estimation across a cluster consisting of $m$ nodes.}
\label{fig:cluster}
\end{figure}

Table \ref{table:circuits} summarizes the circuit size, depth, and width for the various settings of the two algorithms.
Fast phase estimation yields asymptotic improvements in each dimension.
%, achieving in the $m$-cluster case an essentially constant-depth circuit.

\section{Conclusions and Future Work}
We have presented several algorithms for quantum phase estimation based on a basic measurement operation and classical post-processing.
Both our ``information theory" algorithm and our fast phase estimation algorithm depend upon a randomized construction of which measurements to take, and have applications to classical signal processing and quantum phase estimation.
Our fast phase estimation algorithm achieves asymptotic improvements in circuit depth, width, and size over Kitaev's phase estimation, resulting in significant reductions in resource requirements including circuit depth and size, and the number of ancilla qubits.
Remarkably, when using an $m$-node cluster of quantum computers, our algorithm requires essentially constant time.
It is an interesting question for future work to de-randomize these algorithms.

\bibliography{FastPhaseEstNotes}% Produces the bibliography via BibTeX.

\end{document}